\documentclass[twocolumn, tighten, twocolappendix]{aastex631}

\shorttitle{Scaleable redshift inference with \texttt{pop-cosmos}}
\shortauthors{Thorp et al.}

\usepackage{amsmath}	
\usepackage{bm}
\usepackage{booktabs}
\usepackage{threeparttable}
\usepackage[normalem]{ulem}

\DeclareMathOperator{\trace}{Tr}

\newcommand{\add}[1]{{#1}}

\begin{document}

\title{\texttt{pop-cosmos}: Scaleable inference of galaxy properties and redshifts with a data-driven population model}

\correspondingauthor{Stephen Thorp}
\email{stephen.thorp@fysik.su.se}

\author[0009-0005-6323-0457]{Stephen Thorp}
\affiliation{The Oskar Klein Centre, Department of Physics, Stockholm University, AlbaNova University Centre, SE 106 91 Stockholm, Sweden}

\author[0000-0003-4618-3546]{Justin Alsing}
\affiliation{The Oskar Klein Centre, Department of Physics, Stockholm University, AlbaNova University Centre, SE 106 91 Stockholm, Sweden}

\author[0000-0002-2519-584X]{Hiranya V.\ Peiris}
\affiliation{Institute of Astronomy and Kavli Institute for Cosmology, University of Cambridge, Madingley Road, Cambridge CB3 0HA, UK}
\affiliation{The Oskar Klein Centre, Department of Physics, Stockholm University, AlbaNova University Centre, SE 106 91 Stockholm, Sweden}

\author[0000-0003-1943-723X]{Sinan Deger}
\affiliation{Institute of Astronomy and Kavli Institute for Cosmology, University of Cambridge, Madingley Road, Cambridge CB3 0HA, UK}

\author[0000-0002-0041-3783]{Daniel J.\ Mortlock}
\affiliation{Astrophysics Group, Imperial College London, Blackett Laboratory, Prince Consort Road, London, SW7 2AZ, UK}
\affiliation{Department of Mathematics, Imperial College London, London SW7 2AZ, UK}

\author[0000-0002-3962-9274]{Boris Leistedt}
\affiliation{Astrophysics Group, Imperial College London, Blackett Laboratory, Prince Consort Road, London, SW7 2AZ, UK}

\author[0000-0001-6755-1315]{Joel Leja}
\affiliation{Department of Astronomy \& Astrophysics, The Pennsylvania State University, University Park, PA 16802, USA}
\affiliation{Institute for Computational \& Data Sciences, The Pennsylvania State University, University Park, PA 16802, USA}
\affiliation{Institute for Gravitation \& the Cosmos, The Pennsylvania State University, University Park, PA 16802, USA}

\author[0000-0002-4371-0876]{Arthur Loureiro}
\affiliation{The Oskar Klein Centre, Department of Physics, Stockholm University, AlbaNova University Centre, SE 106 91 Stockholm, Sweden}
\affiliation{Astrophysics Group, Imperial College London, Blackett Laboratory, Prince Consort Road, London, SW7 2AZ, UK}



\begin{abstract}
\noindent
We present an efficient Bayesian method for estimating individual photometric redshifts and galaxy properties under a pre-trained population model (\texttt{pop-cosmos}) that was calibrated using purely photometric data. This model specifies a prior distribution over 16 stellar population synthesis (SPS) parameters using a score-based diffusion model, and includes a data model with detailed treatment of nebular emission. We use a GPU-accelerated affine invariant ensemble sampler to achieve fast posterior sampling under this model for 292,300 individual galaxies in the COSMOS2020 catalog, leveraging a neural network emulator (\texttt{Speculator}) to speed up the SPS calculations. We apply both the \texttt{pop-cosmos} population model and a baseline prior inspired by \texttt{Prospector}-$\alpha$, and compare these results to published COSMOS2020 redshift estimates from the widely-used \texttt{EAZY} and \texttt{LePhare} codes. For the $\sim\!12,000$ galaxies with spectroscopic redshifts, we find that \texttt{pop-cosmos} yields redshift estimates that have minimal bias ($\sim10^{-4}$), high accuracy ($\sigma_\text{MAD}=7\times10^{-3}$), and a low outlier rate ($1.6\%$). We show that the \texttt{pop-cosmos} population model generalizes well to galaxies fainter than its $r<25$~mag training set. The sample we have analyzed is $\gtrsim3\times$ larger than has previously been possible via posterior sampling with a full SPS model, with average throughput of 15~GPU-sec per galaxy under the \texttt{pop-cosmos} prior, and 0.6~GPU-sec per galaxy under the \texttt{Prospector} prior. This paves the way for principled modeling of the huge catalogs expected from upcoming Stage IV galaxy surveys. 
\end{abstract}



\section{Introduction}
Photometric redshift (``photo-$z$'') estimates for huge numbers of galaxies will be required for Stage IV large-scale-structure (LSS) imaging surveys, such as the Vera C.\ Rubin Observatory's Legacy Survey of Space and Time \citep[LSST;][]{ivezic19}, {\it Euclid} \citep{laureijs11, euclid24}, and the {Nancy Grace Roman Space Telescope} High-Latitude Imaging Survey \citep{green12, spergel15, dore18, dore19, akeson19, troxel21}, to achieve their core cosmological goals (see \citealp{newman22} for a recent review). For surveys of this kind, photo-$z$ estimates are used to assign galaxies to tomographic redshift bins, with cosmological inference being performed based on the weak-lensing-induced shear of the galaxies in each bin (for a review see e.g. \citealp{bartelmann01, munshi08, kilbinger15, mandelbaum18}). The requirements on photo-$z$ estimators will be extremely stringent for next-generation weak lensing surveys \citep[see e.g.][]{albrecht06, ma06, amara07, mandelbaum08, bordoloi10, bordoloi12, mandelbaum18, desc18, hemmati19, abruzzo19, schmidt20}. Strict control of the rate of ``catastrophic outliers''\footnote{Various definitions exist of the term ``catastrophic outlier''. The most frequently used outlier metric when comparing estimators to spectroscopic redshifts is is $|z-z^\text{spec}|/(1+z^\text{spec})>0.15$. Rubin Observatory requirements are most often stated in terms of the rate of $3\sigma$ outliers; the overall LSST target is for 3$\sigma$ outliers to occur at a rate below 10\% at all redshifts \citep{lsst09, srd, graham23}. Others define catastrophic outliers as only those galaxies where $|z-z^\text{true}|=\mathcal{O}(1)$ \citep{bernstein10, jones24}, or use definitions based on the interquartile range of the sample \citep{graham18}.} will also be needed \citep{sun09, hearin10, bernstein10}. For weak lensing cosmology with first-year LSST data, uncertainties in the mean redshift of each tomographic bin should be smaller than $0.002\times(1+z)$ and the scatter in each bin should be known to within $0.006\times(1+z)$; for the full 10-year analysis, the requirements tighten to $0.001\times(1+z)$ and $0.003\times(1+z)$, respectively \citep{desc18}. Codes that learn a color--redshift relation using self-organizing maps (SOMs; \citealp{wright20, myles21}) have been shown to approach the accuracy required for the first year of LSST, based on Stage III survey data in idealized conditions \citep{newman22}. However, no existing methods have been demonstrated to reach the accuracy requirements of Stage IV surveys when accounting for the additional systematics that will be present in this setting \citep{newman22}.

Considerable interest has been directed towards methods that aim to characterize a survey's redshift distribution, $n(z)$, at the population level. This is usually achieved via hierarchical or forward modeling of entire photometric catalogs \citep[e.g.][]{leistedt16, leistedt19, leistedt23, herbel17, fagioli18, sanchez19, alarcon20, tortorelli18, tortorelli20, tortorelli21, alsing23, alsing24, li23, moser24, autenrieth24, crenshaw24}. By estimating $n(z)$ directly, these approaches sidestep difficulties related to combining individual photo-$z$ probability density functions (see e.g.\ \citealp{leistedt16, malz21, malz22}).

These population-level models can also be used as a prior in inference of the physical properties and redshifts of individual galaxies. The vast majority of photo-$z$ estimation methods operate on a galaxy-by-galaxy basis, but typically use empirical priors that do not correspond to a realistic population distribution. Since the earliest work on photo-$z$ estimation \citep[e.g.][]{baum62, puschell82, koo85, loh86}, many methods have taken an approach of fitting spectral energy distribution (SED) models to observed galaxy photometry. Galaxy models based on stellar population synthesis (SPS; see \citealp{tinsley80, conroy13} for a review) have been critical to these efforts. Commonly they have relied on either small libraries of fixed templates \citep[e.g.][]{puschell82, koo85, gwyn96, kodama99, arnouts99, bolonzella00, babbedge04, ilbert06, ilbert09,  brammer08, kotullafritze09, sawicki12, eriksen19}, or searches over larger pre-constructed grids that span a model's parameter space \citep[e.g.][]{burgarella05, tanaka15, boquien19}. Some of the state-of-the-art amongst SPS-based SED fitters have worked with continuously parametrized models that are constructed ``on the fly'' within a Bayesian posterior sampling routine; examples include \texttt{GalMC} \citep{acquaviva11}; \texttt{BayeSED} \citep{han14}; \texttt{BEAGLE} \citep{chevallard16}; \texttt{BAGPIPES} \citep{carnall18}; \texttt{Prospector} \citep{johnson21}; and \texttt{ProSpect} \citep{robotham20}. This enables inference under more sophisticated models with more free parameters. However, inference is computationally intensive when evaluating these SPS models directly (around 10--25~CPU-hrs per galaxy; \citealp{leja19, leja20, leja22, hahn23, hahn24}), and specifying analytic priors over the model parameters is challenging \citep{leistedt23, alsing23}.

Here, we present an efficient and scaleable method for inferring properties and redshifts of galaxies using the pre-trained galaxy population model \texttt{pop-cosmos}  \citep{alsing24} as a prior. The survey-level forward modeling approach used by \citet{alsing24} builds on advances in hierarchical modeling made in earlier works by \citet{leistedt16, leistedt19, leistedt23} and \citet{alsing23}, and achieves scalability via emulation of SPS with \texttt{Speculator} \citep{alsing20}. This allows \citet{alsing24} to make inferences about the underlying galaxy population probed by a magnitude-limited sample from the Cosmic Evolution Survey (COSMOS; \citealp{scoville07, weaver22}); including the photometric redshift distribution, $n(z)$. Specifically, they use a score-based diffusion model \citep{song20a} to represent a population distribution over 16 SPS parameters, combining this with emulated SPS \citep{alsing20} and a data model that includes detailed treatment of the strength and variance of emission lines \citep{leistedt23}. \citet{alsing24} constrain all model components simultaneously using the 26-band COSMOS2020 photometric data \citep{weaver22}. 

In this work, we use this model as a prior in Bayesian inference of the properties and redshifts of individual galaxies. Combining the \texttt{Speculator} emulator \citep{alsing20} with a GPU-accelerated affine invariant MCMC sampler enables rapid inference for $\sim300,000$ galaxies from the COSMOS2020 catalog \citep{weaver22}. In Section~\ref{sec:methods} we describe our population model and inference methods. In Section~\ref{sec:data} we describe the different data products we make use of. In Section~\ref{sec:results} we demonstrate the performance of our approach on COSMOS data, including a sample of galaxies fainter than our training set. We discuss our results in the context of other methods in Section~\ref{sec:discussion} and summarise our conclusions in Section~\ref{sec:conclusions}.

\section{Methods}
\label{sec:methods}

We infer the SPS properties and redshifts of a sample of galaxies using a Bayesian formalism in which a model for the galaxy population (\S\ref{sec:sps-model}) is combined with the constraints implied by photometric measurements of the individual galaxies (\S\ref{sec:posterior}). Both these components pose considerable numerical challenges (detailed below), but the final result is that we can produce a well-sampled posterior distribution for a galaxy in considerably less than a GPU-minute.

\subsection{Stellar Population Synthesis Model}
\label{sec:sps-model}
We adopt an SPS model for galaxy SEDs that is based on Flexible Stellar Population Synthesis  \citep[\texttt{FSPS};][]{conroy09, conroy10a, conroy10b} with the parametrization used in \texttt{Prospector} \citep{leja17, leja18, leja19_sfh, leja19, johnson21, wang23}. Under this model a galaxy is characterized by 16 SPS parameters: host galaxy stellar mass $\log_{10}(M/M_\odot)$; stellar metallicity, $\log_{10}(Z/Z_\odot)$; star formation rate ratios between seven adjacent lookback time bins, $\Delta\log_{10}(\text{SFR})_{1:6}$ (see \citealp{leja19_sfh}); optical depth of diffuse dust, $\tau_2$; dust law index, $n$; ratio of birth cloud to diffuse attenuation, $\tau_1/\tau_2$; fraction of emission due to AGN, $\ln(f_\text{AGN})$; optical depth of AGN dust torus, $\ln(\tau_\text{AGN})$; gas-phase metallicity, $\log_{10}(Z_\text{gas}/Z_\odot)$; gas ionization, $\log_{10}(U_\text{gas})$; and redshift, $z$. The full list of 16 parameters is denoted $\bm{\vartheta}= (\bm{\varphi}, z)$, with $\bm{\varphi}$ the 15 non-redshift parameters. 

The two-component dust attenuation model in \texttt{FSPS} follows \citet{charlot00}, with the wavelength dependence of the attenuation following a modified \citet{calzetti00} law \citep{noll09}, with the strength of the UV dust bump tied to the slope of the law \citep{kriek13}. The AGN contributions use the \texttt{CLUMPY} models \citep{nenkova08i, nenkova08ii}, as discussed in \citet{leja18}. We use the \citet{chabrier03} stellar initial mass function (IMF). The effect of nebular emission powered by young stars is included based on \texttt{CLOUDY} \citep{ferland13} \add{using the \texttt{FSPS} lookup tables from \citet{byler17}}, with additional corrections to the strengths and variance of emission lines detailed in Section \ref{sec:posterior}. We include a subset of the 44 strongest emission lines from \citet{byler17}, whose strengths have been calibrated following \citet{leistedt23} and \citet{alsing24}. Attenuation by the intergalactic medium is included based on the \citet{madau95} model.

\subsubsection{The \texttt{Prospector} Prior}
\label{sec:prospector-alpha-prior}

Our baseline prior closely follows the \texttt{Prospector}-$\alpha$ model \citep{leja17, leja18, leja19_sfh, leja19}. Hard upper and lower limits on the 16 free parameters are specified in Table 1 of \citet{alsing24}, with these mostly following from \citet{leja19}.

The baseline prior is not uniform within these limits, as it includes several additional physically motivated constraints. We include a prior on redshift proportional to the comoving volume element: $P(z)\propto dV_{\rm co}/dz$ (which can be calculated following e.g.\ \citealp{hogg99}; we use the cosmological parameters from \citealp{planck18})\footnote{\add{This volumetric term is only used for the \texttt{Prospector} prior, and is absent from \texttt{pop-cosmos}. For the redshift range $0.0<z<4.5$ we consider here, the shape of the prior is minimally sensitive to the assumed cosmology.}}. We include a prior term $P(\log_{10}(M/M_\odot)|z)$ based on the double Schechter mass function of \citet{leja20}, with quadratic redshift evolution of the Schechter function parameters. We adopt a prior $P(\log_{10}(Z/Z_\odot)|\log_{10}(M/M_\odot))$ based on the mass--metallicity relation of \citet{gallazzi05}. We apply a soft constraint to enforce that $Z_\text{gas}\gtrsim Z$:
\begin{equation}
    P(Z_\text{gas}|Z) \propto \left[1 + \exp\left(-\frac{\log_{10}(Z_\text{gas}) - \log_{10}(Z)}{0.05}\right)\right]^{-1}.
\end{equation}
This is motivated by the fact that the stellar metallicity averages over stellar populations with a range of ages, and so includes older metal-poor stars \citep{alsing24}. Observations suggest that stellar metallicity is almost always lower than gas-phase metallicity \citep[e.g.][]{gallazzi05, halliday08, lian18, fraser22}. However, this may not always strictly hold if, e.g., recent inflows of metal-poor material contribute significantly to the gas content of a galaxy.

For the prior on the SFR parameters,  $\Delta\log_{10}(\text{SFR})_{1:6}$, we follow \citet{leja19_sfh} by using a Student's-$t$ distribution with two degrees of freedom and a scale parameter of $0.3$. For the diffuse dust component, our baseline prior $P(\tau_2)$ is a truncated normal distribution with its mode at $0.3$, unit variance, and lower truncation at zero \citep{leja19}. Conditional on this, the dust fraction $P(\tau_1/\tau_2|\tau_2)$ has a truncated normal prior with its mode at $\tau_1/\tau_2=1$, a variance of $0.3$, and lower truncation at zero \citep{leja19}. Finally, the dust law index has a normal prior $P(n|\tau_2)$, with mean $-0.095 + 0.111\,\tau_2 + 0.0066\,\tau_2^2$ and standard deviation $0.4$ \citep{alsing23}.

\subsubsection{The \texttt{pop-cosmos} Prior}
\label{sec:pop-cosmos-prior}

The \texttt{pop-cosmos} model \citep{alsing24} specifies a population prior $P(\bm{\vartheta})$ over the 16 SPS parameters $\bm{\vartheta}$ in the form of a score-based diffusion model \citep{song20a}. This is defined by a continuous transform between the target distribution over SPS parameters, $\bm{\vartheta}=\bm{x}(t=0)\sim p_0(\bm{x})$, and a normal base density, $\bm{x}(t=T)\sim p_T(\bm{x})$, with variance $\sigma_T^2$. The forward (noising) diffusion process is defined by a stochastic differential equation (SDE). Specifically, \texttt{pop-cosmos} uses a variance exploding SDE \citep{song20a} given by
\begin{equation}
    \mathrm{d}\bm{x} = g(t)\,\mathrm{d}\bm{w},
\end{equation}
where $\bm{w}$ follows a Wiener process (i.e., Brownian motion). The reverse-time (denoising) transformation can also be expressed as a diffusion process defined by a stochastic differential equation \citep{anderson82, song20a},
\begin{equation}
    \mathrm{d}\bm{x} = -\frac{1}{2}g^2(t)\bm{s}(\bm{x},t)\,\mathrm{d}t + g(t)\,\mathrm{d}\bar{\bm{w}}.
\end{equation}
Here $\bm{s}(\bm{x},t)=\nabla p_t(\bm{x})$ is the score function, and $\bar{\bm{w}}$ follows reverse time Brownian motion. Equivalently, the marginal distributions of $\bm{x}(t)$ at all times can be described by an ordinary differential equation (ODE) of the form \citep{maoutsa20, song20a}
\begin{equation}
    \mathrm{d}\bm{x} = -\frac{1}{2}g^2(t)\bm{s}(\bm{x},t)\,\mathrm{d}t \equiv \tilde{\bm{f}}(\bm{x},t)\,\mathrm{d}t.
    \label{eq:dx_ode}
\end{equation}
In the \texttt{pop-cosmos} model, the score function is represented by a dense neural network with two layers of 128 neurons, and a \texttt{tanh} activation function \citep{alsing24}. In the variance exploding SDE, $g^2(t)=\mathrm{d}\sigma^2(t)/\mathrm{d}t$, with $\sigma^2(t) = \sigma_0(\sigma_T/\sigma_0)^{t/T}$, $\sigma_0=0.01$, $\sigma_T=6$, and $T=1$ \citep{song20a, alsing24}. The ODE representation in Equation \eqref{eq:dx_ode} is equivalent to the continuous normalizing flow model of \citet{grathwohl18} and  \citet{chen18}. 

To incorporate the \texttt{pop-cosmos} prior into a Bayesian analysis, we need a way to evaluate $p_0(\bm{\vartheta})$ for any $\bm{\vartheta}$. \citet{chen18} show that for $\bm{x}(t)$ with dynamics obeying Equation \eqref{eq:dx_ode}, the log probability density of $\bm{x}(t)$ also follows a differential equation:
\begin{equation}
    \frac{\partial \ln p_t(\bm{x}(t))}{\partial t} = -\trace(\nabla \tilde{\bm{f}}(\bm{x},t)) = -\nabla \cdot \tilde{\bm{f}}(\bm{x},t).
    \label{eq:dp_ode}
\end{equation}
Following \citet{grathwohl18}, Equations \eqref{eq:dx_ode} and \eqref{eq:dp_ode} can be solved simultaneously by integrating from $t=0$ to $t=T$, so 
\begin{equation}
    \bm{x}(t=T) - \bm{x}(t=0) = \int^T_0 \tilde{\bm{f}}(\bm{x}(t),t)\,\mathrm{d}t
    \label{eq:xsolve}
\end{equation}
and
\begin{multline}
    \ln p_T(\bm{x}(t=T)) - \ln p_0(\bm{x}(t=0))  \\= -\int^T_0 \trace(\nabla \tilde{\bm{f}}(\bm{x},t))\,\mathrm{d}t,
    \label{eq:psolve}
\end{multline}
where the initial value $\bm{x}(t=0)=\bm{\vartheta}$ is known. The quantity of interest $p_0(\bm{\vartheta})$ can be computed by rearranging Equation \eqref{eq:psolve} (c.f.\ equation 39 in \citealp{song20a}), and substituting in the solution for $\bm{x}(t=T)$ from Equation \eqref{eq:xsolve} to obtain
\begin{equation}
    \ln p_0(\bm{\vartheta}) = \ln p_T(\bm{x}(t=T)) + \int^T_0\trace(\nabla \tilde{\bm{f}}(\bm{x},t))\,\mathrm{d}t.
    \label{eq:logprob}
\end{equation}
In practice, we use a lower integration limit of $\epsilon=10^{-5}$, rather than exactly zero \citep[see][appendix C]{song20a}. We perform the integration using the \citet{dormand80} adaptive Runge--Kutta routine, following \citet{song20a} and using the implementation by \citet{chen18}. We use an absolute (\texttt{atol}) and relative (\texttt{rtol}) error tolerance of $10^{-4}$, since we find this offers an acceptable trade-off between reliability and computational cost (see Appendix \ref{sec:numerical}). The trace of the Jacobian in the integrand can be found using Hutchinson trace estimator \citep{skilling89, hutchinson89}, as per \citet{grathwohl18}, or can be computed directly. The Hutchinson estimator gives
\begin{equation}
    \trace(\nabla \tilde{\bm{f}}(\bm{x},t)) \approx \mathbb{E}[\bm{\epsilon}^\top \nabla\tilde{\bm{f}}(\bm{x},t) \bm{\epsilon}],
    \label{eq:hutch}
\end{equation}
where and $\bm{\epsilon}$ is drawn from a distribution with zero mean vector and identity covariance matrix \citep{grathwohl18, song20a}. A Rademacher distribution of the form $P(\epsilon)=0.5[\delta(\epsilon-1) + \delta(\epsilon+1)]$, where $\delta$ is the Dirac delta function, satisfies these criteria \citep{hutchinson89}, as would iid.\ normal variates $\bm{\epsilon}\sim N(\bm{0},\mathbf{I})$ (\citealp{silver94}; see \citealp{avron11, adams18} for further discussion). In either approach, the gradient terms in $\nabla \tilde{\bm{f}}(\bm{x},t)$ can be computed using automatic differentiation. The Hutchinson estimator is advantageous in the sense that vector--Jacobian products can be computed more efficiently than the full Jacobian using automatic differentiation. However, we find that for the low-dimensional Jacobian involved in our model there is not a significant advantage to using the approximation (see Appendix \ref{sec:numerical}). 

\subsection{Single Galaxy Posterior Distribution}
\label{sec:posterior}

Our primary aim here is, for a single galaxy, to evaluate 
$P(\bm{\vartheta} | \hat{\mathbf{f}})$, the posterior distribution of its SPS parameters and redshift, $\bm{\vartheta} = (\bm{\varphi}, z)$, conditional on the measured fluxes, $\hat{\mathbf{f}} = (\hat{\mathrm{f}}_1, \hat{\mathrm{f}}_2, \ldots, \hat{\mathrm{f}}_B)$, where $B$ is the number of bands.
Under the assumption that the photometric measurements in the different bands are independent, 
\begin{equation}
    P(\bm{\vartheta}|\hat{\mathbf{f}}) \propto P(\bm{\vartheta})\times\prod_{b=1}^{B}P(\hat{\mathrm{f}}_b|\bm{\vartheta}),
\end{equation}
where the prior $P(\bm{\vartheta})$ comes from either the baseline \texttt{Prospector}-$\alpha$ model (\S\ref{sec:prospector-alpha-prior}) or the \texttt{pop-cosmos} diffusion model (\S\ref{sec:pop-cosmos-prior}). After defining the photometric likelihood (\S\ref{sec:photlik}), the inference then proceeds via a two-step process in which maximum a posteriori (MAP) estimates (\S\ref{sec:map}) are used as the starting point for full posterior sampling (\S\ref{sec:mcmc}).

\subsubsection{Photometric Likelihood}
\label{sec:photlik}

To evaluate the likelihood for the data given $\bm{\vartheta}$, we first synthesize model fluxes $\mathbf{f}^\text{SPS}(\bm{\vartheta})$ using the \texttt{Speculator} emulator (\citealp{alsing20}; emulating the SPS model described in \S\ref{sec:sps-model}). 

After computing the flux contributions $\mathbf{f}^\text{EM}_b(\bm{\vartheta})$ of the 44 emission lines to the flux in a band $b$, the total model flux in that band is given by $\mathrm{f}_b(\bm{\vartheta}) = {\alpha}^\text{ZP}_b [\mathrm{f}_b^\text{SPS}(\bm{\vartheta}) + \bm{\beta}^\text{EM}\cdot\mathbf{f}^\text{EM}_b(\bm{\vartheta})]$, where $\alpha_b^\text{ZP}$ is a zero-point correction for band $b$, and $\bm{\beta}^\text{EM}$ is a vector of emission line strength corrections (c.f.\ \citealp{alsing24} eq.\ 2). The uncertainty on the emission line contribution to band $b$ will be given by $\sigma_{\text{EM},b} = \alpha^\text{ZP}_b\bm{\gamma}^\text{EM}\circ(\bm{\beta}^\text{EM}+1)\cdot\mathbf{f}^\text{EM}_b$, where $\bm{\gamma}^\text{EM}$ is a vector of emission line variances. We also include an error floor term $\sigma_{\text{floor},b}=\alpha^\text{ZP}_b \times y_b \times \mathrm{f}_b^\text{SPS}(\bm{\vartheta})$, where all $y_b$ are $\leq0.1$ and fixed to the values inferred by \citet{leistedt23}. For the baseline prior, we use the COSMOS zero-point corrections $\bm{\alpha}^\text{ZP}$, emission line strength corrections $\bm{\beta}^\text{EM}$, and emission line strength variances $\bm{\gamma}^\text{EM}$ estimated by \citet{leistedt23}. For the \texttt{pop-cosmos} prior, we use the updated values of these calibration parameters from \citet{alsing24}. 

As in \citet{leistedt23}, we take the likelihood to be a heavy-tailed Student's-$t$ distribution with two degrees of freedom, so that  for the $b$th band
\begin{equation}
P(\hat{\mathrm{f}}_b|\bm{\vartheta}) = \mathcal{T}_2\big(\hat{\mathrm{f}}_b|\mathrm{f}_b(\bm{\vartheta}), \hat{\sigma}_b^2 + \sigma_{\text{EM},b}^2 + \sigma_{\text{floor},b}^2 \big),
\end{equation}
where
\begin{equation}
    \mathcal{T}_2(x|l, s^2) \propto \frac{1}{s}\left(1 + \frac{(x - l)^2}{2s^2}\right)^{-3/2}.
\end{equation}

\subsubsection{Maximum a Posteriori Estimation}
\label{sec:map}
As a starting point, we draw a large sample of galaxies from the baseline prior or \texttt{pop-cosmos} diffusion model and compute synthetic photometry for these. Since this is purely an initialization step, our results are insensitive to the exact number of prior draws taken here. In practice, we use 6.4 million draws. For each real galaxy, we find the prior draw with the highest likelihood, giving us a pseudo-MAP estimate of $\bm{\vartheta}$ for that galaxy. We then refine these estimates via stochastic gradient descent, using the \texttt{Adam} optimizer \citep{kingma14} with the negative log-posterior as our objective function.

\subsubsection{Posterior Sampling}
\label{sec:mcmc}
Using our MAP estimates as a starting point, we obtain posterior samples for each galaxy using an affine-invariant ensemble Markov chain Monte Carlo (MCMC) sampler \citep{goodman10}. We use a GPU-accelerated implementation\footnote{\url{https://github.com/justinalsing/affine}} where the log probability calls are parallelized over batches of galaxies. For the baseline prior, we handle batches of 4,000 galaxies simultaneously, and use an ensemble of 512 walkers to explore the posterior distribution. Evaluating the \texttt{pop-cosmos} prior is more computationally intensive due to the ODE solver calls, so we use a batch size of 1,000 galaxies, and an ensemble of 512 walkers. We run chains for 500 iterations, discarding the first 250. We thin the chains by a factor of 50, leaving a final set of $2,560$ posterior draws per galaxy. Due to this thinning procedure the effective sample size is $\sim\!2\times10^3$, giving us an excellent numerical representation of $P(\bm{\vartheta} | \hat{\mathbf{f}})$.

\section{Data}
\label{sec:data}

We apply our inference method (\S\ref{sec:methods}) to data from the Cosmic Evolution Survey (COSMOS; \citealp{scoville07}). All data products correspond to Version 2.1 of the COSMOS2020 catalog \citep{weaver22} for consistency with \citet{leistedt23} and \citet{alsing24}. Spectroscopic data are taken from a variety of companion surveys, which we detail in \S\ref{sec:cosmos-spec}. We also make use of photometric redshifts included in the COSMOS2020 catalog (detailed in \S\ref{sec:cosmos-photoz}).

\subsection{Photometric Data}
\label{sec:cosmos-photometry}
We use the COSMOS2020 photometric galaxy catalog \citep{weaver22}. As in \citet{alsing24}, we select the data from the \texttt{Farmer} photometric pipeline, which is based on a forward modeling approach \citep{weaver23}. We use the same set of $B = 26$ bands as \citet{leistedt23} and \citet{alsing24} -- this includes $u$-band data from the Canada--France--Hawaii Telescope MegaPrime/MegaCam; optical ($grizy$) data from Subaru Hyper Suprime-Cam; near IR ($YJHK_s$) data from UltraVISTA; mid IR data from {Spitzer} IRAC; and narrow band data from Subaru Suprime-Cam. As in \citet{alsing24} we apply a magnitude cut of $r<25$, giving a sample of 140,745 galaxies. We also impose a requirement that all observed fluxes and all signal-to-noise ($\mathrm{S/N}$) ratios for an object are $>0$. We also omit objects which are flagged as failing the star--galaxy separation cut applied by \citet{weaver22}. Their cut is made using a mixture of morphological information, and stellar template fits using \texttt{LePhare} \citep{arnouts99, ilbert06, ilbert09}. The $r<25$ photometric catalog includes 1,256 galaxies that are flagged as having a significant X-ray detection (likely due to the presence of an active galactic nucleus) in {Chandra} data \citep{civano16, marchesi16}.

We also perform fits to an expanded sample that is cut based on {Spitzer} IRAC Channel 1 magnitude: $\texttt{irac1}<26$. For this sample, we also require $\text{flux}>0$ and $\mathrm{S/N}>0$ in all bands. This leaves a total of 292,300 galaxies that pass the \citet{weaver22} star--galaxy separation criterion. Of these, 152,188 have $r>25$, meaning they are outside the training sample to which the \texttt{pop-cosmos} model was calibrated. There are 1,545 galaxies in this sample with a significant X-ray detection. This \texttt{irac1} magnitude limit was used by \citet{weaver23_smf} to study the galaxy stellar mass function out to high-$z$, and will be studied in depth by Deger et al. (in prep.) using \texttt{pop-cosmos}.

\subsection{Spectroscopic Data}
\label{sec:cosmos-spec}

Of the 140,745 galaxies in our $r<25$ photometric sample, there is a subset of 12,014 with accurate redshifts \citep{weaver22}, obtained from a variety of spectroscopic surveys that have targeted the COSMOS field. Around half of these (6,567) are from the $z$COSMOS survey \citep{lilly07}, with 2,881 from the DEIMOS 10k catalog \citep{hasinger18}, and 1,849 from the Complete Calibration of the Color-Redshift Relation Survey (C3R2; \citealp{masters17, masters19, stanford21}). We include further spectroscopic redshifts from MUSE (337 galaxies; \citealp{rosani20}), the Subaru Fiber Multi-Object Spectrograph (FMOS; 303 galaxies; \citealp{kashino19}), the VIMOS VLT Deep Survey (VVDS, 56 galaxies; \citealp{lefevre05}), and the VIMOS Ultra Deep Survey (VUDS, 21 galaxies; \citealp{lefevre15}). Of this spectroscopic subset, 501 sources in total are flagged as having a significant X-ray detection; $\sim\!4\%$ of galaxies.

The $\texttt{irac1}<26$ sample includes 12,309 galaxies with spectroscopic redshifts: 297 of these have $r>25$; and 526 of these are X-ray sources. The redshifts for these fainter galaxies come from DEIMOS (138 galaxies; \citealp{hasinger18}), C3R2 (66 galaxies; \citealp{masters17, masters19, stanford21}), MUSE (50 galaxeis; \citealp{rosani20}), FMOS (40 galaxies; \citealp{kashino19}), and VUDS (3 galaxies; \citealp{lefevre15}).

A large subset of these galaxies were analysed with an SPS-based hierarchical Bayesian model by \citet{leistedt23}.

\subsection{Photometric Redshifts}
\label{sec:cosmos-photoz}
As well as spectroscopic redshifts, photometric redshifts are publicly available for COSMOS2020 \citep{weaver22}. These come from two codes: \texttt{LePhare} \citep{arnouts99, ilbert06, ilbert09} and \texttt{EAZY} \citep{brammer08}. 

The \texttt{LePhare} photometric redshift estimates are based on the fitting of galaxy SED templates to the observed photometry. \citet{weaver22} use a library of 33 galaxy templates (the template set used by \citealp{polletta07, onodera12}; from the SPS models of \citealp{silva98, bruzual03}), along with several possible dust attenuation laws \citep{prevot84, calzetti00}, and a set of AGN templates \citep{salvato09, salvato11}. A set of stellar templates \citep{bohlin95, pickles98, chabrier00, morley12, morley14, baraffe15} is also included in the \texttt{LePhare} fits done by \citet{weaver22} -- COSMOS sources where a stellar SED is a better fit than any galaxy template are flagged. In the catalog studied here and in \citet{alsing24}, these objects are omitted.

The \texttt{EAZY} photometric redshift code is also based on a finite set of galaxy templates (originally from \citealp{bruzual03}). \citet{weaver22} use \texttt{EAZY} with an updated set of 17 templates generated using \texttt{FSPS} \citep{conroy09, conroy10a, conroy10b}. Galaxy SEDs are fit using a linear combination of these 17 \texttt{FSPS} templates. A library of theoretical stellar templates \citep{allard12} are also included in \texttt{EAZY}, although \citet{weaver22} do not use these for performing star--galaxy separation. For both \texttt{EAZY} and \texttt{LePhare}, \citet{weaver22} use an iterative procedure to estimate zero-point corrections based on the subsample with spectroscopic redshifts. This is done by fitting the templates to the spectroscopic subsample, with the redshift fixed to the known spec-$z$, and then applying a correction based on the mean residual ($\text{observed mag}-\text{model mag}$) in each band. This process is repeated until the correction stabilizes between iterations \citep{ilbert06}.

The \texttt{pop-cosmos} model \citep{alsing24} we use in this paper is built on an emulator for \texttt{FSPS} \citep[\texttt{Speculator};][]{alsing20}, making the \texttt{EAZY} results from \citet{weaver22} a useful comparison. In effect, our \texttt{pop-cosmos} model is leveraging a continuously-parameterized version of the underlying model that \texttt{EAZY} was built from (see also discussion in \citealp{leistedt23}).

\begin{figure*}
    \centering
    \includegraphics[width=\linewidth]{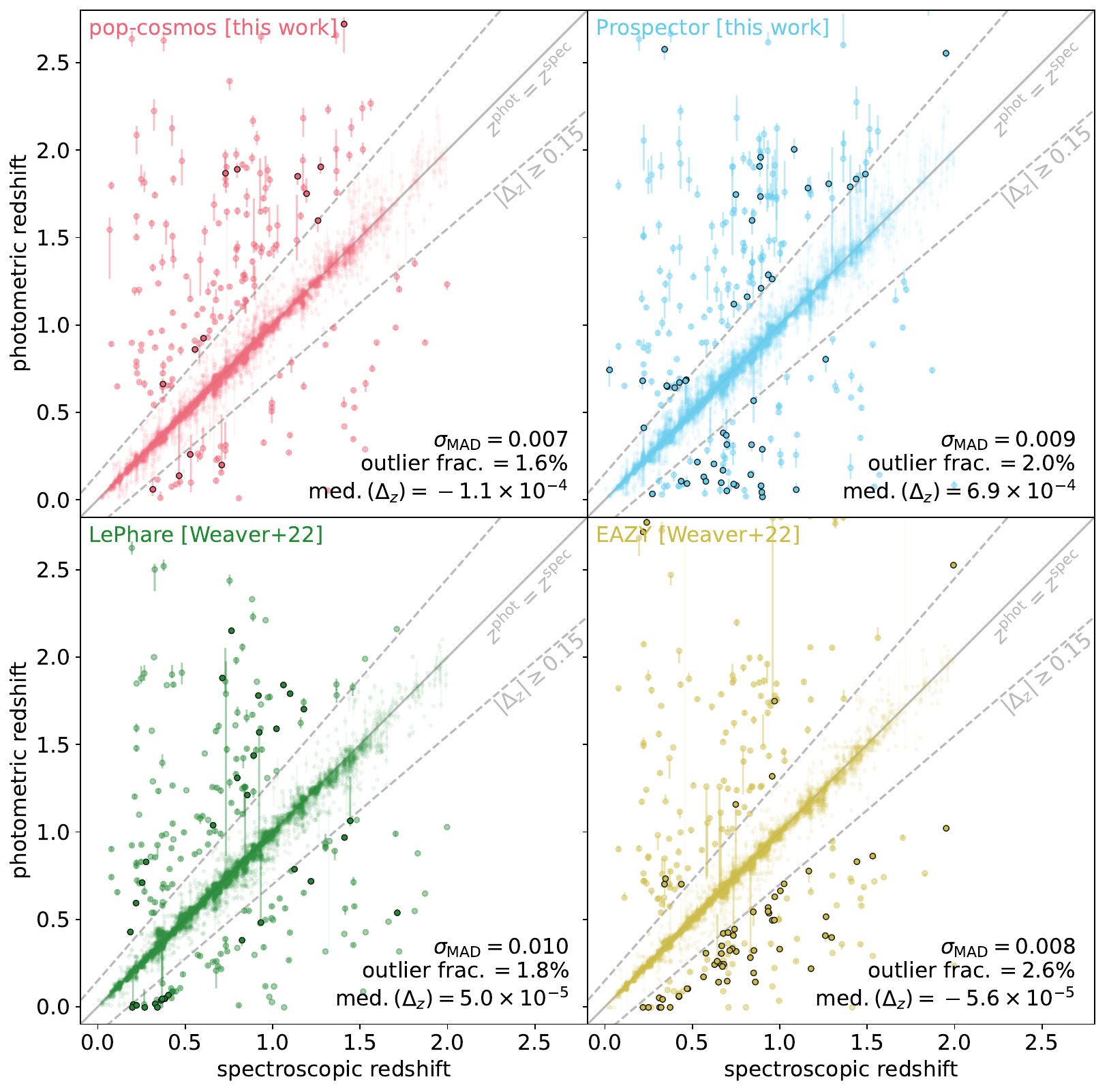}
    \caption{Photometric redshift vs.\ spectroscopic redshift for $\sim\!12,000$ COSMOS galaxies using \texttt{pop-cosmos} (\S\ref{sec:pop-cosmos-prior}; \citealp{alsing24}), \texttt{Prospector}-$\alpha$ (\S\ref{sec:prospector-alpha-prior}; \citealp{leja17, leja18, leja19_sfh, leja19}), \texttt{LePhare} \citep{arnouts99, ilbert06, ilbert09}, and \texttt{EAZY} \citep{brammer08}. Results for \texttt{pop-cosmos} and \texttt{Prospector} are from this work; plotted points are posterior medians and 68\% credible intervals. Results for \texttt{LePhare} and \texttt{EAZY} are from \citet{weaver22}; plotted points are posterior medians for \texttt{LePhare} and MAP estimates for \texttt{EAZY}. Error bars show the reported 68\% confidence levels for both codes. The threshold of $|\Delta_z|>0.15$ is indicated on all panels with dashed grey lines, where $\Delta_z=(z-z^\text{spec})/(1+z^\text{spec})$. Galaxies where $|\Delta_z|>0.15$ are indicated with more solid markers. \add{Galaxies that are only outlying for one of the codes are shown with black circles.}}
    \label{fig:photo-vs-spec}
\end{figure*}

\section{Results}
\label{sec:results}
In this Section, we present the results of applying the methodology detailed in \S\ref{sec:methods} to the data described in \S\ref{sec:data}. In \S\ref{sec:spec-results}, we present results on the subset of COSMOS2020 where spectroscopic redshifts are available. In \S\ref{sec:phot-results}, we present photometric redshifts for a larger sample of COSMOS2020 galaxies with $r<25$~mag. Finally in \S\ref{sec:irac-results}, we give our results for galaxies with $r>25$~mag and $\texttt{irac1}<26$ -- i.e.\ galaxies fainter and/or redder than the \texttt{pop-cosmos} training sample.

\subsection{Spectroscopic Sample}
\label{sec:spec-results}
The spectroscopic redshift of a galaxy, $z^\text{spec}$, gives us something close to a ground truth with which we can compare\footnote{However, spectroscopic surveys do suffer from a low but non-zero ($\lesssim1\%$; \citealp{newman22}) rate of catastrophic failures even amongst objects flagged as having a secure redshift. Failures are typically the result of misidentified lines, or superpositions of multiple galaxies (see e.g.\ \citealp{lefevre05, lilly07, newman13, baldry14, cunha14}).}. Figure \ref{fig:photo-vs-spec} plots a direct comparison of photometric and spectroscopic redshifts for the 12,014 galaxies in the spectroscopic sample. For \texttt{pop-cosmos} and \texttt{Prospector}, the photo-$z$s plotted are the posterior medians and 68\% credible intervals. For the \texttt{LePhare} results from \citet{weaver22}, we plot the posterior median (\texttt{lp\_zBEST} in the COSMOS2020 catalog) and 68\% credible interval (which are obtained under a nominally flat prior). For objects where there is an X-ray detection ($\texttt{lp\_type}=2$ in the COSMOS2020 catalog), we use the \texttt{LePhare} redshift corresponding to the best fitting AGN template (\texttt{lp\_zq} in the COSMOS2020 catalog). For the \citet{weaver22} \texttt{EAZY} results the plotted points are MAP estimates, and the error bars are 68\% credible intervals. For 444 \texttt{EAZY} objects, the photo-$z$ MAP is not contained within the 68\% credible interval reported by \citet{weaver22}. We set the error bar to zero for these objects.

\begin{figure}
    \centering
    \includegraphics[width=\linewidth]{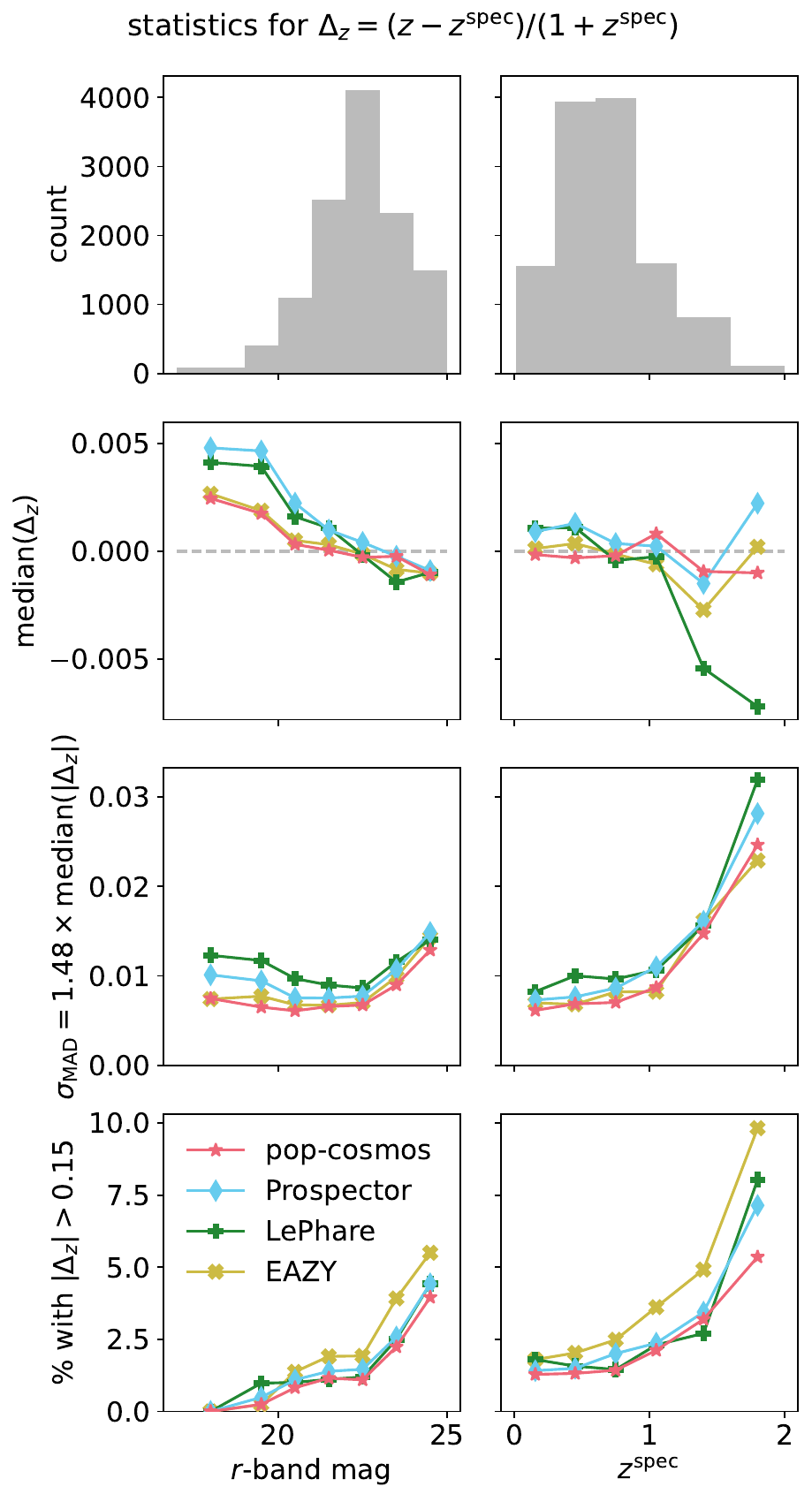}
    \caption{Statistics comparing the different photometric redshift estimates to spectroscopic redshifts for the $\sim\!12,000$ COSMOS galaxies (c.f.\ also fig.\ 2 in \citealp{leistedt23}). Left column shows the statistics in bins of $r$-band magnitude; right column shows bins of $z^\text{spec}$. All statistics are computed for $\Delta_z=(z - z^\text{spec})/(1+z^\text{spec})$.}
    \label{fig:photo-z-metrics}
\end{figure}

In Figure \ref{fig:photo-z-metrics}, we compare \texttt{pop-cosmos} to \texttt{Prospector}, \texttt{LePhare}, and \texttt{EAZY} using the same set of photo-$z$ metrics as \citet{leistedt23}. All of these metrics are computed based on the quantity $\Delta_z=(z-z^\text{spec})/(1+z^\text{spec})$; i.e.\ the absolute error made by each photo-$z$ estimator, scaled by $1/(1+z^\text{spec})$. For \texttt{pop-cosmos} and \texttt{Prospector} here, we use the posterior medians. In the left column of Figure \ref{fig:photo-z-metrics} we show the metrics binned by $r$-band magnitude, with the right hand column being binned by $z^\text{spec}$. The second row shows the median of $\Delta_z$, measuring the bias of each estimator. For \texttt{pop-cosmos}, the bias is generally the smallest of the four methods, and is fairly constant with magnitude and redshift. The third row of Figure \ref{fig:photo-z-metrics} shows the robust standard deviation of $\Delta_z$ for the four methods, given by $\sigma_\text{MAD}=1.48\times\mathrm{median(}|\Delta_z|)$. This gives a sense of the level of scatter of each estimator about the truth, with \texttt{pop-cosmos} and \texttt{EAZY} being the the lowest for all magnitude and redshift bins. This result also compares favourably to the fully hierarchical model used by \citet{leistedt23}, which fixed many aspects of the population prior to \texttt{Prospector}-like settings. Finally, the fourth row of Figure \ref{fig:photo-z-metrics} shows the fraction of galaxies with $|\Delta_z|>0.15$, measuring the rate of outliers. Overall, \texttt{pop-cosmos} has the lowest outlier fraction in almost all bins, and shows the least steep increase in this quantity towards fainter magnitudes or higher $z^\text{spec}$. The total number of outliers for each method are listed in Table \ref{tab:outliers}, and the total outlier fractions are printed on Figure \ref{fig:photo-vs-spec}. All of the methods show some degree of bias towards overestimating $z$ for the outliers; the numbers of outliers where $\Delta_z>0.15$ are also tabulated in Table \ref{tab:outliers} along with the fraction of all outliers that are overestimates. \add{The fraction of outliers that are overestimates is highest for \texttt{pop-cosmos} (75\% of its outliers), although the total number of outliers is lowest, and 92.8\% of these are also outliers for at least one other estimator, suggesting the inference from the photometry is not unreasonable. Unique outliers are highlighted with black circles on Figure \ref{fig:photo-vs-spec}.} \add{Additional details are given in Table \ref{tab:delz}.}

\begin{table}
    \centering
    \caption{Outlier counts ($|\Delta_z|>0.15$) for the photo-$z$ estimators tested. Reported are total outlier counts (out of 12,014 galaxies), number of overestimates ($\Delta_z>0.15$), and number of AGN that are outliers.}
    \label{tab:outliers}
    \begin{tabular}{c c c c c c}
        \toprule
        & \multicolumn{3}{c}{\# of outliers} & \multicolumn{2}{c}{\% of all outliers} \\
        \cmidrule(lr){2-4} \cmidrule(l){5-6}
        Estimator & All & Overest.\ & AGN & Overest.\ & AGN\\
        \midrule
        \texttt{pop-cosmos} & 195 & 147 & 59 & 75\% & 30\% \\
        \texttt{Prospector} & 235 & 156 & 77 & 66\% & 33\% \\
        \texttt{LePhare} & 216 & 130 & 58 & 60\% & 27\%\\
        \texttt{EAZY} & 316 & 206 & 88 & 65\% & 28\%\\
        \bottomrule
    \end{tabular}
\end{table}

For all of the methods, X-ray selected AGN are outliers by the $|\Delta_z|>0.15$ metric at a relatively high rate. The number of outliers for each method that are X-ray sources (flagged by \citealp{weaver22} using {Chandra} data; \citealp{civano16, marchesi16}) is included in Table \ref{tab:outliers}. For all methods, around 30\% of outliers coincide with an X-ray source. For \texttt{pop-cosmos}, $11.8\%$ of the 501 X-ray sources in the spectroscopic sample are outliers. For \texttt{LePhare}, using the redshift estimates based on the \citet{salvato09, salvato11} AGN templates, the fraction is similar. The relatively high fraction of outliers amongst X-ray detected sources is unsurprising; optical variability of AGN is known to introduce additional scatter into photo-$z$ estimates \citep{babbedge04, simm15}.

\begin{figure}
    \centering
    \includegraphics[width=\linewidth]{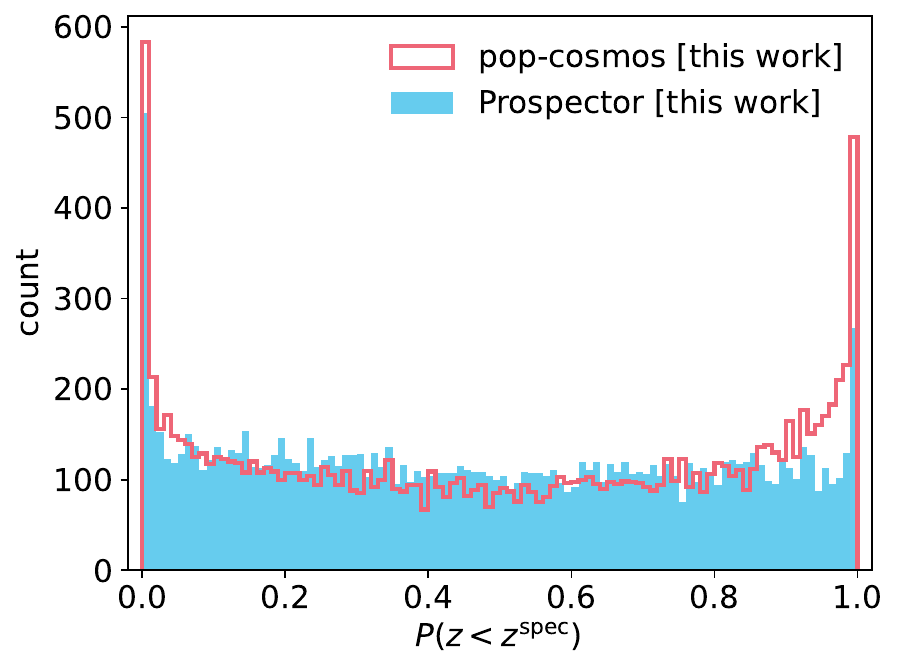}
    \caption{Probability integral transform (PIT) test for the redshift posteriors under the \texttt{pop-cosmos} and \texttt{Prospector} models. For each galaxy in the COSMOS spectroscopic sample, we estimate the posterior probability that $z<z^\text{spec}$.}
    \label{fig:pit}
\end{figure}

As a further check of the individual redshift posteriors estimated under the \texttt{pop-cosmos} model, we perform a test based on the probability integral transform (PIT; e.g.\ \citealp{bordoloi10, wittman16, freeman17, tanaka18, schmidt20, newman22}). This is shown in Figure \ref{fig:pit}. For each galaxy in the COSMOS spectroscopic sample, we estimate the posterior probability that $z<z^\text{spec}$. This is given by
\begin{equation}
P(z<z^\text{spec}|\hat{\mathbf{f}}) = \int_0^{z^\text{spec}} P(z|\hat{\mathbf{f}})\,\mathrm{d}z,
\end{equation}
where
\begin{equation}
    P(z|\hat{\mathbf{f}}) = \int P(\bm{\vartheta}|\hat{\mathbf{f}})\,\mathrm{d}\bm{\varphi}
\end{equation}
is a marginalization over $\bm{\varphi}$. If the posterior under our model was perfectly calibrated and the population model itself was correct, the distribution of $P(z<z^\text{spec}|\hat{\mathbf{f}})$ would be close to uniform. Figure \ref{fig:pit} shows that this is not the case. However, the distribution is fairly symmetric about $0.5$, suggesting there is no strong bias in our results. The ``bathtub'' shape indicates that the posteriors under the \texttt{pop-cosmos} prior are systematically slightly overconfident. 

Also shown in Figure \ref{fig:pit} are the results for the baseline \texttt{Prospector} prior described in Section \ref{sec:prospector-alpha-prior}. These results are closer to uniform, reflective of the broader and more conservative prior. The posteriors under the \texttt{Prospector} prior are wider and frequently have longer tails; this gives better coverage in the PIT test. However, the point estimates (posterior medians) are less accurate, and have a higher outlier rate based on the metrics in Figure \ref{fig:photo-z-metrics}. As illustrated by \citet{schmidt20}, an estimator with perfect frequentist calibration can have arbitrarily bad individual redshift performance; the downstream interpretation of the posteriors will dictate the relative importance of these qualities. For both \texttt{pop-cosmos} and \texttt{Prospector}, the rate of extreme overestimates (the left hand spike; caused by $>99\%$ of the posterior samples being above $z^\text{spec}$) is higher than the rate of extreme underestimates (the right hand spike; coming from $>99\%$ of posterior samples being $<z^\text{spec}$). This is unsurprising given that: (a) redshift is constrained to be positive; and (b) the COSMOS spectroscopic sample is not representative, leaning towards the lower end of the redshift ranges permitted by either of our priors. We note that the majority of the codes tested on simulated data by \citet{schmidt20} show departures from uniformity in the PIT test at least as extreme as the results in our Figure \ref{fig:pit}. Post-hoc recalibration of photometric redshift posteriors has been proposed \citep{bordoloi10, zhao21, dey21, dey22}, but we do not explore this here.

For \texttt{EAZY}, $48.3\%$ of galaxies have $z^\text{spec}$ within the reported 68\% credible interval. For \texttt{LePhare}, $z^\text{spec}$ is within the reported 68\% credible interval for $51.7\%$ of galaxies. For \texttt{Prospector} and \texttt{pop-cosmos}, the equivalent numbers are respectively $62.8\%$ and $55.0\%$.

\begin{figure*}
    \centering
    \includegraphics[width=0.5\linewidth]{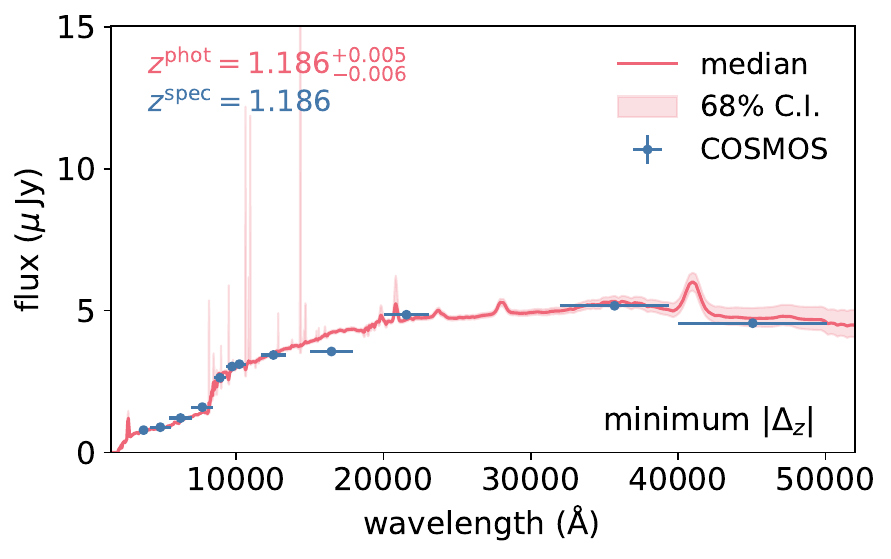}%
    \includegraphics[width=0.5\linewidth]{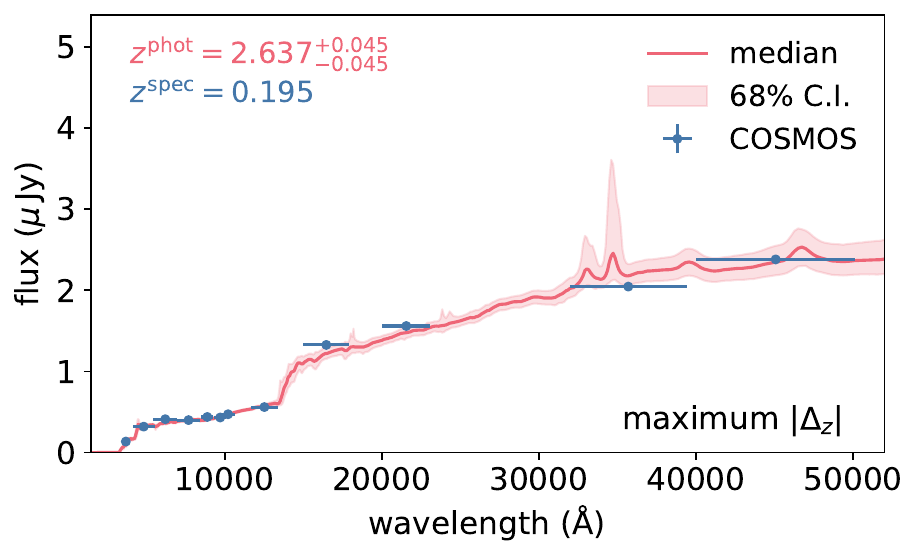}
    \includegraphics[width=0.5\linewidth]{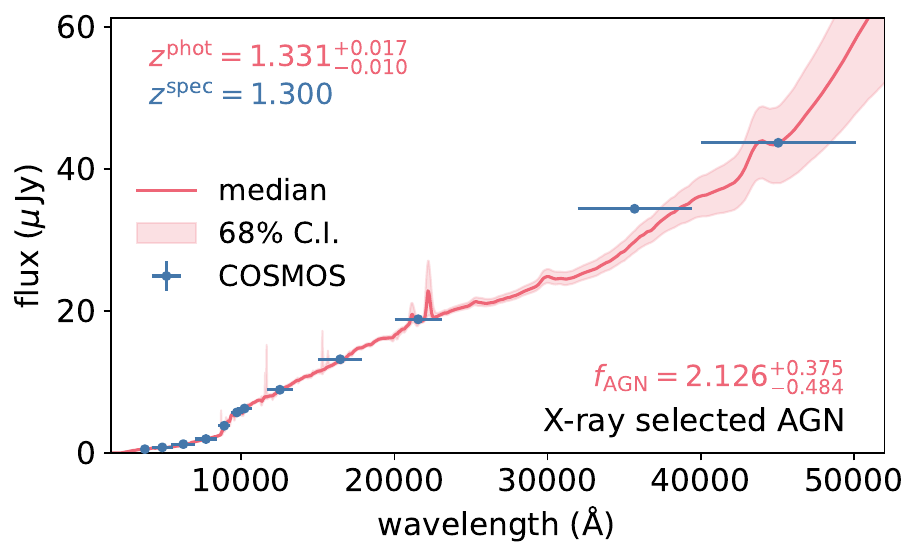}
    \includegraphics[width=0.5\linewidth]{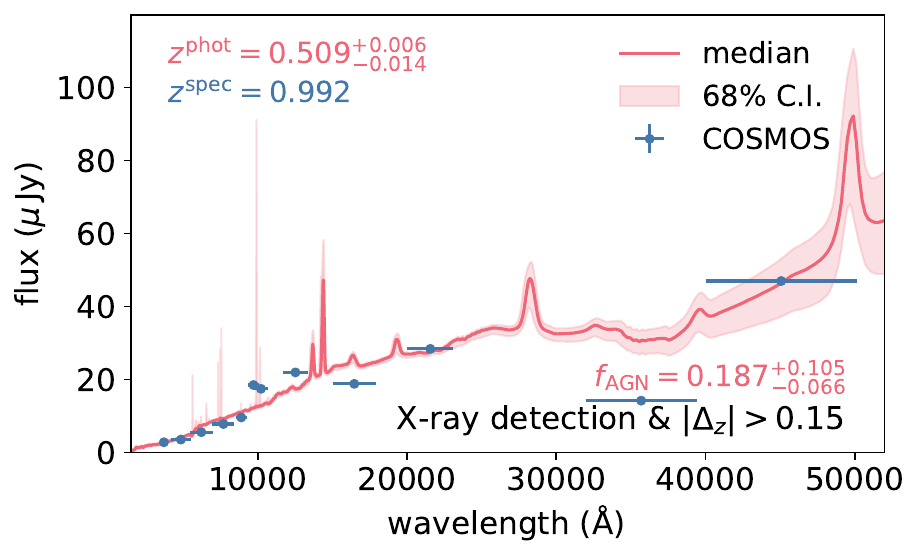}
    \caption{Example SED posteriors for individual galaxies under the \texttt{pop-cosmos} prior. \add{\textbf{Top left:}} the galaxy with minimum $|\Delta_z|$; i.e.\ the posterior median closest to $z^\text{spec}$. \add{\textbf{Top right:}} the galaxy with maximum $|\Delta_z|$; i.e.\ with $z^\text{phot}$ posterior median furthest from $z^\text{spec}$. \add{\textbf{Bottom left:}} an X-ray selected AGN for which we find $f_\text{AGN}>2$; i.e.\ an SED with twice as much AGN flux as galaxy flux. \add{\textbf{Bottom right:} a galaxy with a significant X-ray detection that is a large outlier with $|\Delta_z|>0.15$.} Fluxes are in microjanskys. Corresponding redshift posteriors are in Figure \ref{fig:redshiftposteriors}. Horizontal bars on the COSMOS data indicate the FWHM of the passbands.} 
    \label{fig:sedfits}
\end{figure*}

In Figure \ref{fig:sedfits}, we show the SED posterior for several individual galaxies in the spectroscopic sample. The SEDs are computed using \texttt{FSPS} \citep{conroy09, conroy10a, conroy10b} via the \texttt{python-fsps} bindings \citep{pythonfsps}. For each MCMC sample of the parameters $\bm{\vartheta}$, we compute the rest-frame SED for those parameters. We then redshift these into the observer frame using the redshift of each MCMC sample, and interpolate these observer-frame SEDs onto a common wavelength grid. The plotted SED is then the pointwise median and 68\% credible interval at each wavelength. Note that marginalizing over the uncertain redshift leads to a smoothing out or broadening of emission lines. For this visualization, nebular emission is applied using the default \texttt{CLOUDY} prescription \citep{ferland13, byler17} in FSPS. We overplot the COSMOS broadband photometry on the SED, as well as the reported $z^\text{spec}$ and a summary (median and 68\% credible interval) of the redshift posterior. Photometric points are placed at the filter central wavelength, with the horizontal error bar corresponding to the full width at half maximum (FWHM) of the transmission curve for that filter. Central wavelengths and FWHMs are tabulated in \citet{weaver22}.

The top \add{left} panel of Figure \ref{fig:sedfits} shows the galaxy where the \texttt{pop-cosmos} redshift posterior is closest to the spectroscopic redshift (i.e.\ the minimum $|\Delta_z|$). The posterior distribution over the model SED shows a strong Balmer break feature, clearly visible in the data as a flux deficit blueward of the $z$-band. 

The \add{top right} panel of Figure \ref{fig:sedfits} shows the galaxy where \texttt{pop-cosmos} has the largest $|\Delta_z|$, with a estimate of $z=2.637^{+0.045}_{-0.045}$ vs.\ a reported spectroscopic redshift of $0.195$. Visually the fit to the data is very good, with the high redshift estimate being connected to a strong flux excess in the $H$-band relative to the $J$-band. In the model SED, this is well described by a Balmer break between these two bands. This particular galaxy also has the largest $|\Delta_z|$ for \texttt{Prospector} and \texttt{LePhare}. With \texttt{Prospector} we estimate $z=2.635^{+0.052}_{-0.068}$, whilst \citet{weaver22} report a \texttt{LePhare} redshift of $z=2.623^{+0.037}_{-0.038}$. This galaxy is not the largest outlier in $|\Delta_z|$ for \texttt{EAZY}, but it has a similar redshift estimate of $z=2.680_{-0.104}^{+0.010}$. This object (COSMOS-319050; coordinates 10h\,02m\,00.65s, +02$^\circ$\,21$'$\,3.1$''$) has a quality flag of 4 in the C3R2 catalog indicating the highest level of confidence in the assigned redshift \citep{masters17}. \add{The spectrum of the galaxy is included in Appendix \ref{sec:spectra}.} Deeper investigation of this galaxy, and other similar objects where the spectroscopic redshift and SED fit are incompatible, is needed to diagnose the source of the disagreement.

The bottom \add{left} panel of Figure \ref{fig:sedfits} shows an X-ray selected AGN for which we estimate a high fraction of AGN contribution in the SED ($f_\text{AGN}>2$ with $>50$\% posterior probability). The fit to the data in this case is visually very good, and the posterior median redshift is close to the reported $z^\text{spec}$. \add{By contrast, in the bottom right panel we show an X-ray detected galaxy that is a large outlier for \texttt{pop-cosmos}. In this case the inferred AGN contribution is low but non-negligible, and the fit is visually less good; the model SED fails to reproduce the relatively bright $J$-band or faint $H$-band and \texttt{irac1} fluxes. The spectrum for this galaxy is discussed in Appendix \ref{sec:spectra}.}

\begin{figure*}
    \centering
    \includegraphics[width=\linewidth]{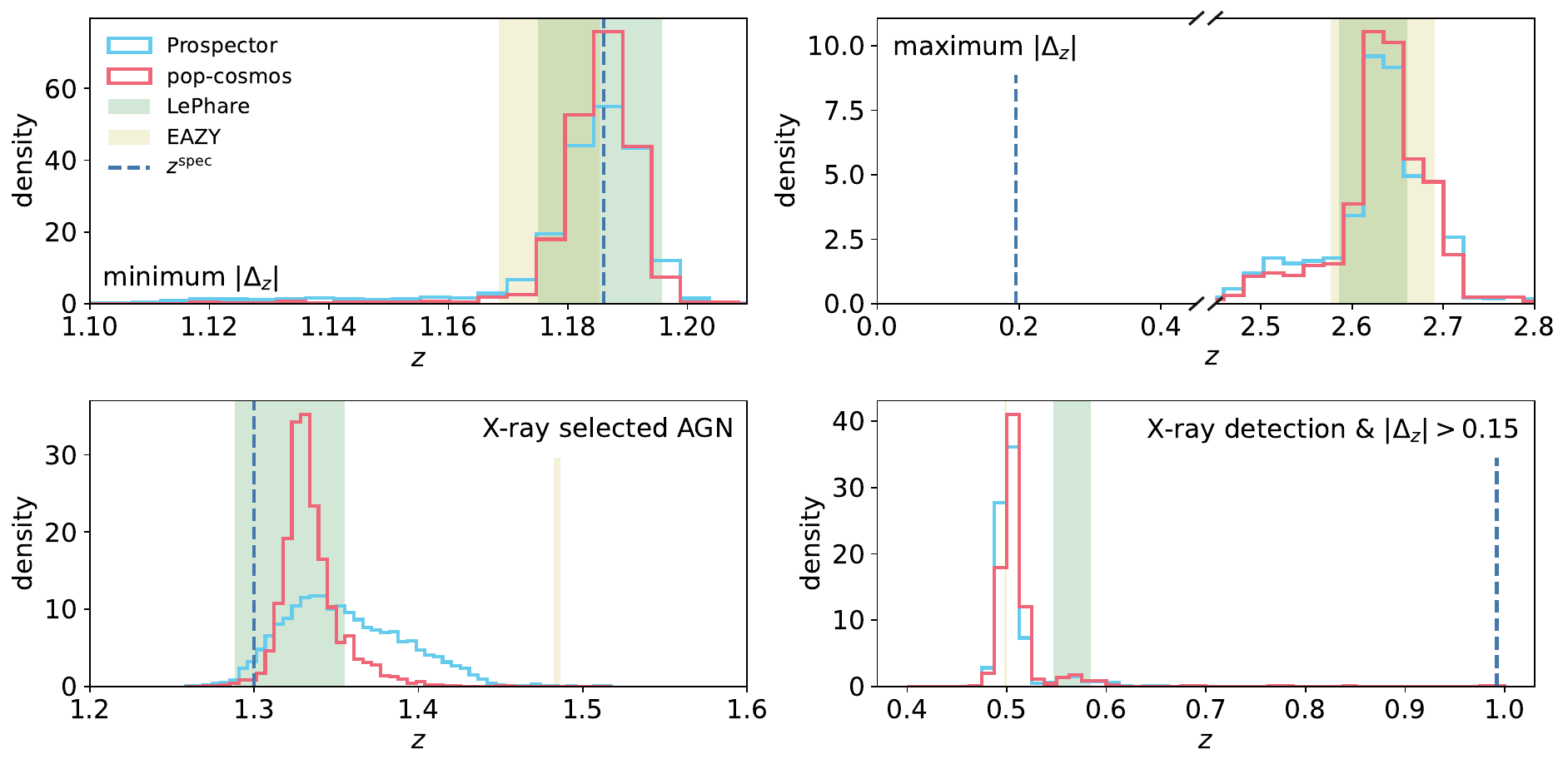}
    \caption{Redshift posteriors for the \add{four} galaxies shown in Figure \ref{fig:sedfits}. Posteriors for \texttt{pop-cosmos} and \texttt{Prospector} are shown as histograms; credible intervals for \texttt{LePhare} and \texttt{EAZY} are shaded regions. Spectroscopic redshift is shown as a dashed line.}
    \label{fig:redshiftposteriors}
\end{figure*}

Figure \ref{fig:redshiftposteriors} shows the marginal posterior distributions over redshift for the three example galaxies shown in Figure \ref{fig:sedfits}. The \add{top left} panel corresponds to the galaxy where \texttt{pop-cosmos} has the smallest $|\Delta_z|$. The \texttt{pop-cosmos} and \texttt{Prospector} posteriors are very similar here, and are consistent with \texttt{LePhare} and \texttt{EAZY}. The left tail on the posterior from \texttt{Prospector} is heavier, reflective of the broader prior. The \add{top right} panel shows the galaxy where \texttt{pop-cosmos} has the largest $|\Delta_z|$. All of the models estimate $z\approx2.6$, meaning they are all very far from the reported $z^\text{spec}=0.195$.

\begin{figure*}
    \centering
    \includegraphics[width=\linewidth]{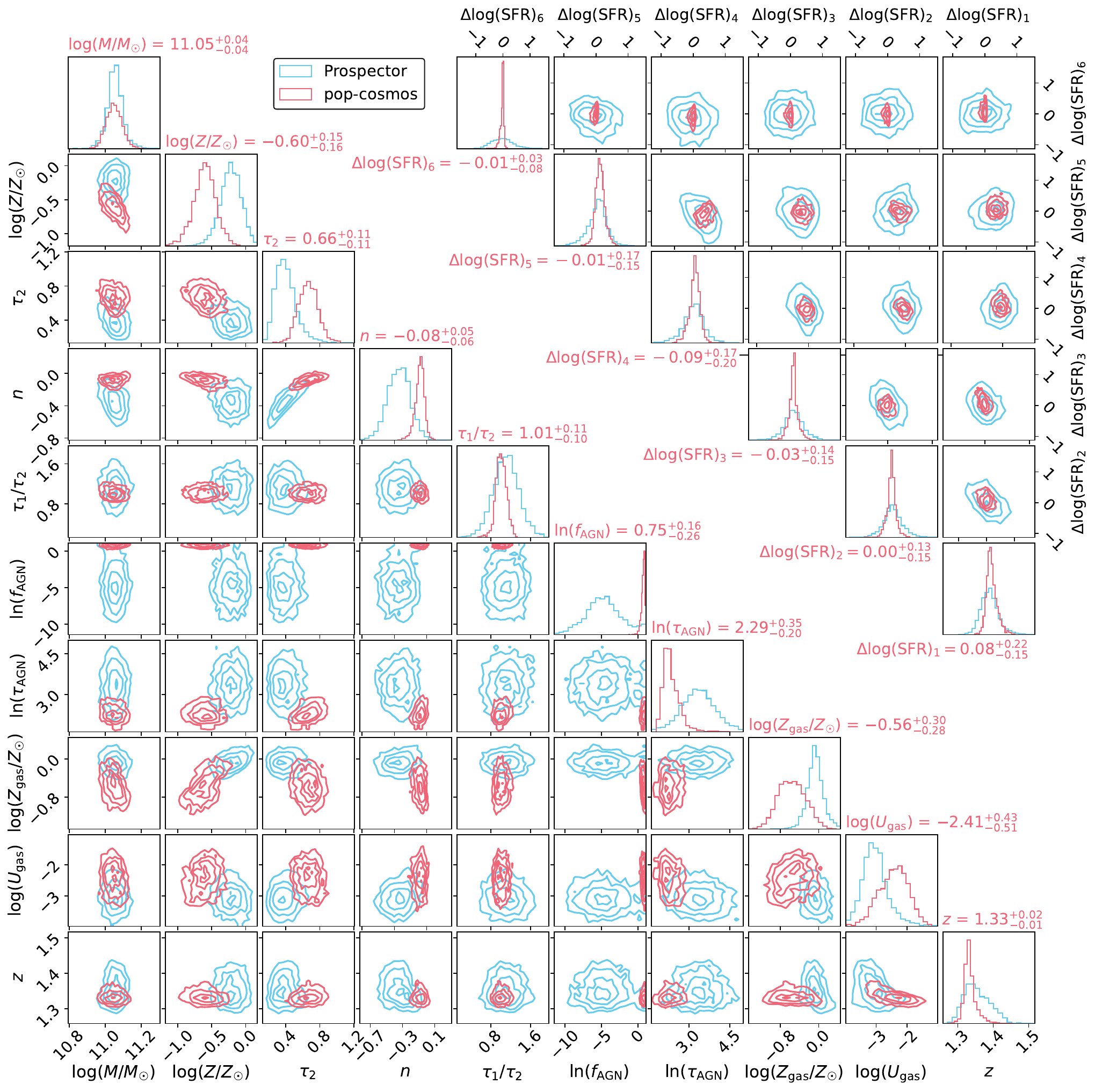}
    \caption{Full 16 dimensional posterior distributions of $\bm{\vartheta}$ under the \texttt{pop-cosmos} \add{(red)} and \texttt{Prospector} \add{(blue)} priors for the X-ray selected AGN (lower \add{left} panel of Figures \ref{fig:sedfits} and \ref{fig:redshiftposteriors}).}
    \label{fig:fullposteriors}
\end{figure*}

The lower \add{left} panel of Figure \ref{fig:redshiftposteriors} shows the posteriors for the X-ray selected AGN. Here, we see that \texttt{pop-cosmos}, \texttt{Prospector}, and \texttt{LePhare} are in broad agreement, with \texttt{pop-cosmos} having the tightest posterior. In this example we find $P(z<z^\text{spec})=0.012$, placing it into the second lowest bin of the PIT plot shown in Figure \ref{fig:pit}. This is indicative of how a galaxy that would not be classed as an outlier by the $|\Delta_z|>0.15$ criterion can still fall towards the outer edges of the $P(z<z^\text{spec})$ distribution. The \texttt{EAZY} redshift estimate is very discrepant in this example; likely because \citet{weaver22} do not explicitly include any AGN models in their \texttt{EAZY} template set. In contrast, \add{\texttt{Prospector}, \texttt{pop-cosmos}, and \texttt{LePhare} include AGN templates as described in Sections \ref{sec:sps-model} and \ref{sec:cosmos-photoz}}. For this galaxy, we inferred a very high AGN contribution to the SED under the \texttt{pop-cosmos} prior ($f_\text{AGN}=2.13^{+0.38}_{-0.48}$). Under the \texttt{Prospector} prior, the estimated AGN fraction spans the full prior range. Figure \ref{fig:fullposteriors} shows the full 16-parameter posteriors on $\bm{\vartheta}$ under the \texttt{pop-cosmos} and \texttt{Prospector} priors for this example.

\begin{figure}
    \centering
    \includegraphics[width=\linewidth]{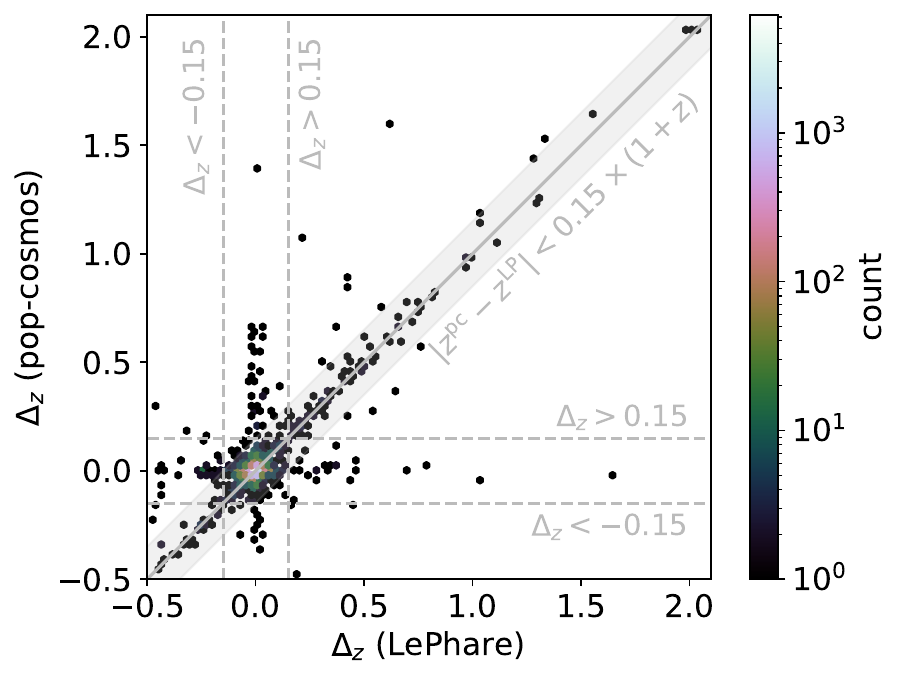}
    \caption{Comparison of $\Delta_z=(z-z^\text{spec})/(1+z^\text{spec})$ for \texttt{pop-cosmos} and \texttt{LePhare}. \add{Each cell is colored based on the number of galaxies (out of the spectroscopic sample) it contains. The grey shaded region indicates where \texttt{LePhare} and \texttt{pop-cosmos} agree to within $0.15\times(1+z)$.}}
    \label{fig:delz-delz}
\end{figure}

\add{The redshift posterior in the lower right panel of Figure \ref{fig:redshiftposteriors} corresponds to the exemplar X-ray detected galaxy which has a large $|\Delta_z|$ for \texttt{pop-cosmos}. The agreement between \texttt{pop-cosmos}, \texttt{Prospector}, and \texttt{EAZY} is very close here. The estimate from \texttt{LePhare} is slightly higher, and aligns with a minor mode in the \texttt{pop-cosmos} and \texttt{Prospector} posteriors. This galaxy is one of the largest redshift underestimates for \texttt{pop-cosmos} and has $\Delta_z\lesssim0.4$ for all four estimators. For \texttt{pop-cosmos}, 15 of 59 AGN outliers are underestimates. Thus the underestimate rate for AGN is the same as for all galaxies (c.f.\ Table \ref{tab:outliers} and Table \ref{tab:delz}), and so this example does not appear to be indicative of a broader trend.}

We include in Figure \ref{fig:delz-delz} a density plot comparing the $\Delta_z$ values for \texttt{pop-cosmos} and \texttt{LePhare}. The axes are divided into different regions by the dashed lines depending on whether one, both, or neither of the codes estimated a redshift that is strongly discrepant from $z^\text{spec}$. For example, the upper right square contains galaxies where both codes strongly overestimated $z$ relative to the reported $z^\text{spec}$. The highest density is in the ``ideal'' region, where the two methods agree with each other and the reported $z^\text{spec}$. Out of the 12,014 galaxies in the sample, 11,747 (97.8\%) fall in the ideal region. Outside of this, the vertical band between the dashed lines corresponds to galaxies that are outliers for \texttt{pop-cosmos} but not \texttt{LePhare} (51 galaxies), and the horizontal band contains the opposite situation (72 galaxies). The diagonal shaded region in Figure \ref{fig:delz-delz} highlights the region where \texttt{pop-cosmos} and \texttt{LePhare} disagree with the reported $z^\text{spec}$ in a similar way. There are 120 galaxies in this region (including the example\add{s} shown in the \add{right hand} panel\add{s} of Figure\add{s} \ref{fig:sedfits} and \ref{fig:redshiftposteriors}); these are interesting objects for further investigation. For 102 of these objects all four photo-$z$ estimates are consistent to within $0.15\times(1+z)$. \add{Various related quantities are summarized in Table \ref{tab:delz}. There are only 14 galaxies for which \texttt{pop-cosmos} is the sole outlier with respect to the reported $z^\text{spec}$. Overall, an accurate and concordant redshift estimate, with $|\Delta_z|<0.15$ for all four estimators, is achieved for 96.3\% of the COSMOS spectroscopic sample. Of the 438 galaxies that are outlying for any estimator, 53\% are outlying for at least two, and 29\% are outlying for all four. For these galaxies, investigating the causes of mismatches (e.g.\ break confusion in the SED fits or spectra, misidentified lines, or spurious cross-matches between catalogs) would be an interesting avenue for further work.}

\begin{table}
    \centering
    \caption{Similarity of outliers with $|\Delta_z|>0.15$ between the four photo-$z$ estimators (c.f.\ Figure \ref{fig:delz-delz}).}
    \label{tab:delz}
    \begin{tabular}{l c r}
        \toprule
        Criterion & Estimator & Count \\
        \midrule
        $\Delta_z < -0.15$ & Any & 192\\
        & \texttt{pop-cosmos} only & 8\\
        & \texttt{Prospector} only & 32\\
        & \texttt{LePhare} only & 31\\
        & \texttt{EAZY} only & 53\\
        & All & 29\\
        $\Delta_z < -0.15$ \& X-ray detected & Any & 52\\
        & All & 5\\
        $\Delta_z > 0.15$ & Any & 280\\
        & \texttt{pop-cosmos} only & 11\\
        & \texttt{Prospector} only & 29\\
        & \texttt{LePhare} only & 17\\
        & \texttt{EAZY} only & 74\\
        & Only one & 131\\
        & Any two & 28 \\
        & Any three & 32\\
        & All & 89\\
        $|\Delta_z|>0.15$ & None & 11576\\
        & Any & 438\\
        & \texttt{pop-cosmos} only & 14\\
        & \texttt{Prospector} only & 51\\
        & \texttt{LePhare} only & 34\\
        & \texttt{EAZY} only & 107\\
        & Only one & 206\\
        & Any two & 66\\
        & Any three & 40\\
        & All & 126\\
        $|z - z^\texttt{pc}| < 0.15\times(1+z^\text{spec})$ & \texttt{Prospector} & 143\\
        & \texttt{LePhare} & 120\\
        & \texttt{EAZY} & 130\\
        & All & 102\\
        \bottomrule
    \end{tabular}
\end{table}

\subsection{Photometric Sample}
\label{sec:phot-results}
\begin{figure}
    \centering
    \includegraphics[width=\linewidth]{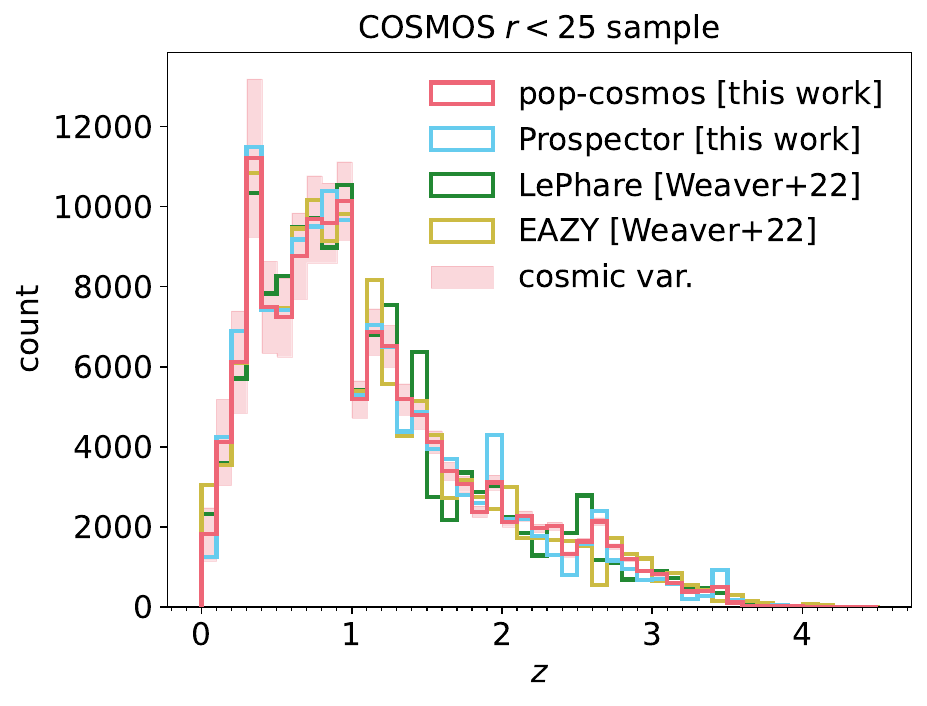}
    \caption{Photometric redshift distribution for the full $r<25$ COSMOS sample from the four estimators. \add{Posterior medians are used for \texttt{pop-cosmos}, \texttt{Prospector}, and \texttt{LePhare}; maximum a-posteriori (MAP) estimates are used for \texttt{EAZY}.}}
    \label{fig:nz-map}
\end{figure}

For the full photometric sample of 140,745 galaxies, we can compare our posterior median photo-$z$ estimates to the published COSMOS photo-$z$s from \texttt{LePhare} and \texttt{EAZY} \citep{weaver22}. Figure \ref{fig:nz-map} shows histograms of the photo-$z$ point estimates from the four methods. The agreement is generally very good, with many of the prominent features being shared by the three distributions -- e.g.\ the sudden drop at $z\approx1$, and the spike at $0.3\leq z<0.4$. On each histogram bin for \texttt{pop-cosmos} we plot a shaded region showing the expected effect of cosmic variance on a COSMOS-sized field ($84'\times84'$), calculated using the recipe in \citet{moster11}. The spike at $0.3\leq z<0.4$ could plausibly be explained by cosmic variance. In the tail of the distribution at $z>1$, \texttt{pop-cosmos} generally shows less spikes and discontinuities than \texttt{LePhare}, \texttt{EAZY}, and \texttt{Prospector} -- particularly around $1.5<z<2$ and $2.5<z<3$.

\newpage
\subsection{Spitzer IRAC Selected Sample}
\label{sec:irac-results}
\begin{figure}
    \centering
    \includegraphics[width=\linewidth]{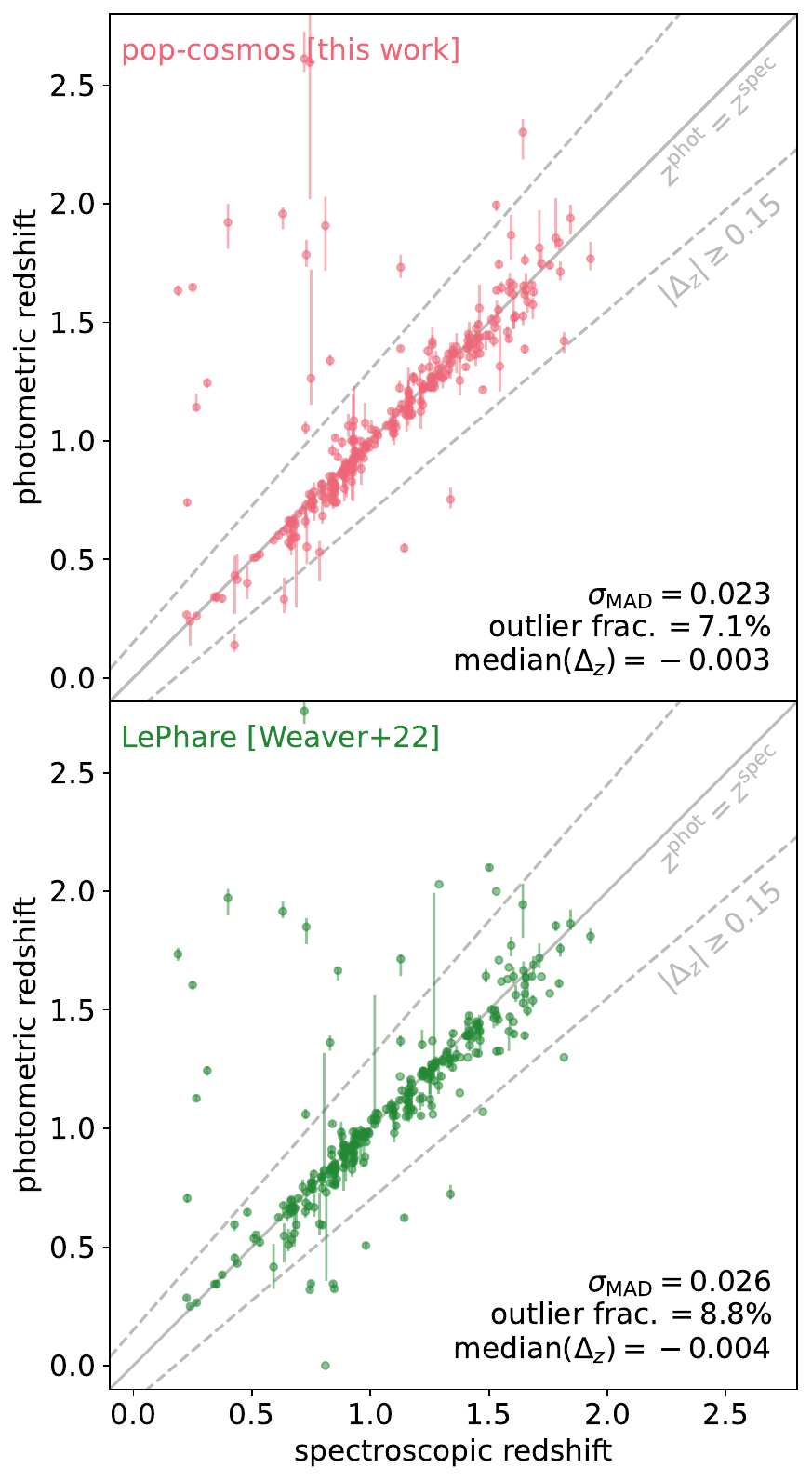}
    \caption{Photometric vs.\ spectroscopic redshifts for the 297 out-of-training-sample galaxies that have $r>25$ and $\texttt{irac1}<26$. Results are shown for \texttt{pop-cosmos} and \texttt{LePhare}.}
    \label{fig:photo-vs-spec-irac}
\end{figure}

We also perform fits under the \texttt{pop-cosmos} prior to an expanded set of galaxies with $\texttt{irac1}<26$, $\text{flux}>0$ in all bands, and $\mathrm{S/N}>0$ in all bands. This sample is around double the size of our fiducial $r<25$ sample; all galaxies with $\texttt{irac1}<26$ and $r>25$ are outside of the training set used in \citet{alsing24}. In Figure \ref{fig:photo-vs-spec-irac}, we show the \texttt{pop-cosmos} photo-$z$ estimates for the 297 galaxies with $r>25$ that have available spectroscopic redshifts. Despite these galaxies being faint in the $r$-band (with $25<r\lesssim27$), the agreement between spectroscopic and photometric redshift is excellent. We find a median $\Delta_z$ of $-0.003$, and $\sigma_\text{MAD}=0.023$ -- comparable to the results on the $r<25$ sample (\S\ref{sec:spec-results}). The outlier rate is higher ($7.1$\%) than for the $r<25$ sample, but lower than \texttt{LePhare} ($8.8\%$), and less than half that of \texttt{EAZY} ($16.5$\%). The median $\Delta_z$ for \texttt{LePhare} and \texttt{EAZY} is $-0.004$ and $-0.006$, respectively; their $\sigma_\text{MAD}=1.48\times\text{median}(|\Delta_z|)$ scores are $0.026$ and $0.027$. We also include \texttt{LePhare} in Figure \ref{fig:photo-vs-spec-irac} as a point of comparison, since this is the fiducial model used by \citet{weaver23_smf} in their analysis of an $\texttt{irac26}<26$ sample.

\begin{figure}
    \centering
    \includegraphics[width=\linewidth]{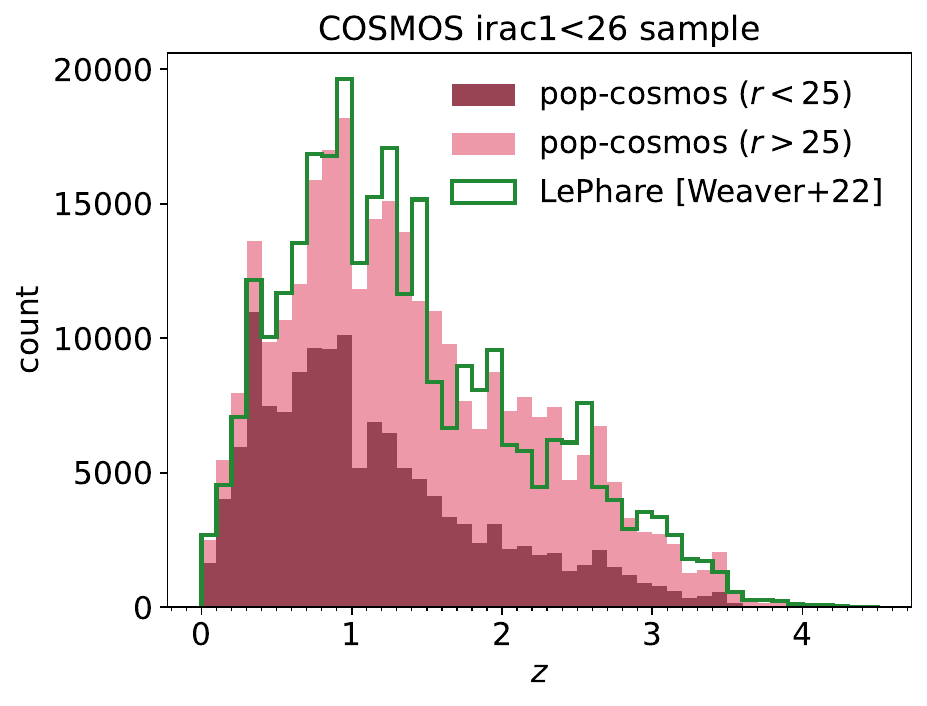}
    \caption{Redshift posterior medians for COSMOS galaxies with $\texttt{irac1}<26$. The histogram bars for \texttt{pop-cosmos} are stacked; the fraction of each bar within the $r<25$ cut that \citet{alsing24} used on the training set is shown in maroon, with the fraction with $r>25$ shown in lighter pink. The total height of the bar corresponds to the total number of galaxies in that redshift bin (irrespective of $r$-band magnitude). The green histogram shows \texttt{LePhare} results from \citet{weaver22}.}
    \label{fig:nz-irac}
\end{figure}

In Figure \ref{fig:nz-irac}, we show the distribution of photometric redshift posterior medians obtained under the \texttt{pop-cosmos} prior for the full $\texttt{irac1}<26$ sample (292,300 galaxies). Around half of these galaxies are fainter than the $r<25$ training sample used by \citet{alsing24}; these are distributed across the full redshift range, but make up a larger fractional contribution at $z\gtrsim1$. At $z<1$, around 67\% of galaxies have $r<25$. For $1<z<3$ only $\sim\!37\%$ have $r<25$, and the fraction drops to $\sim\!7\%$ for galaxies with an estimated $z>4$. Although there are bin-to-bin disagreements between \texttt{pop-cosmos} and the \texttt{LePhare} redshifts from \citet{weaver22}, the overall shapes of the distributions are very consistent. Features such as the cleft at $1.0\leq z<1.1$, the spike at $0.3\leq z<0.4$, and the peak around $2.5\leq z<2.8$ are not washed out in the more complete $\texttt{irac1}<26$ sample. On a galaxy by galaxy basis, the agreement between \texttt{pop-cosmos} and \texttt{LePhare} is generally very good; the two estimators agree to within $0.15\times(1+z)$ for 93\% of galaxies.

\section{Discussion}
\label{sec:discussion}
In Section \ref{sec:context}, we provide further discussion and context relating to current progress in the modeling of galaxy populations using SPS. In Section \ref{sec:scaling}, we provide details about computational scalability (see also Appendix \ref{sec:numerical}) in the context of other approaches.

\subsection{Galaxy Population Modeling with SPS}
\label{sec:context}
The SPS parametrization we have used (see \S\ref{sec:sps-model}) is highly flexible; motivated by the state-of-the-art \texttt{Prospector} models \citep{leja17, leja18, leja19_sfh, leja19, johnson21, wang23}. The recent \texttt{Prospector}-$\beta$ update by \citet{wang23} combines this parametrization with a set of informative priors motivated by previous galaxy evolution studies. This is in contrast to the \texttt{Prospector}-$\alpha$ model \citep{leja17, leja18, leja19_sfh, leja19}, our baseline for comparison (see \S\ref{sec:prospector-alpha-prior}), which uses broader uniform priors that span the full range of physically acceptable parameter values. The approach we take with the \texttt{pop-cosmos} prior introduced by \citet{alsing24} is complementary to the route taken by \citet{wang23}, which is motivated by breaking degeneracies in photometric redshift estimation for high-$z$ galaxies observed with {JWST} \citep[e.g.][]{wang24}. Rather than constraining the wider parameter space by leveraging empirically derived relationships to break degeneracies, we instead use a population distribution that was calibrated to reproduce the color and magnitude distributions of a deep dataset with excellent wavelength coverage (COSMOS; \citealp{weaver22}).

Recent work by \citet{tortorelli24} has compared the \texttt{Prospector}-$\beta$ model from \citet{wang23} to a series of variations that are possible within FSPS. They simulate galaxy colors using the \texttt{Prospector}-$\beta$ model and associated priors, then assign these to tomographic redshift bins using a self-organizing map (SOM) trained on Dark Energy Spectroscopic Instrument (DESI) spectra and KiDS--VIKING photometry \citep{masters15, mccullough24}. They repeat this exercise with various modifications to the SPS model, and test for a change in the true redshift distribution of the galaxies assigned to each tomographic bin based on their colors. This analysis approach is different to the one we have used here (where we are directly fitting an SPS model to galaxy photometry), and to the full population modeling approach taken by \citet{alsing24}. 

\citet{tortorelli24} find that galaxy colors and tomographic bin assignments in their approach are insensitive to the treatment of late stages of stellar evolution -- e.g.\ populations of Wolf--Rayet or (post) asymptotic giant branch stars -- or the inclusion of intergalactic medium absorption. They also show that galaxy colors are not strongly influenced by the use of a \citet{salpeter55} or \citet{kroupa01} IMF in place of that of \citet{chabrier03}. They see a significant impact on the SED when switching to a fixed slope \citet{calzetti00} attenuation law or a Milky Way-like \citet{cardelli89} extinction law, in place of the more general attenuation law \citep{noll09, kriek13} and two-population dust model \citep{charlot00} used in \texttt{Prospector} and \texttt{pop-cosmos}. Their results also indicate that galaxy colors and the color--redshift relation are noticeably impacted by restricting the allowed gas physics; e.g.\ by removing nebular emission, fixing $\log(U_\text{gas})$, or requiring $Z_\text{gas}=Z$. This strongly motivates the flexible approach taken here, and by \citet{leistedt23, alsing24}. The importance of flexible handling of nebular emission was also recently noted by \citet{csornyei21}.

Our approach to modeling emission line strengths and variances, introduced by \citet{leistedt23}, is a generalization of the standard \texttt{CLOUDY} \citep{ferland13} treatment used in \texttt{FSPS} \citep{conroy09, conroy10a, conroy10b, byler17} and \texttt{Prospector} (although an analytical marginalization over line strengths is discussed by \citealp{johnson21}). Moreover, this model is calibrated directly on photometric data, as an alternative to using bright and nearby spectroscopic samples \citep[e.g.][]{khederlarian24}.

\subsection{Computational Efficiency}
\label{sec:scaling}
By performing our inference using an affine-invariant sampler with batched posterior evaluation, we are able to obtain posterior samples for huge numbers of galaxies in a relatively short time. We find that 256,000 prior evaluations are possible in 25~s on a single GPU; an effective rate of $0.1$~ms per prior evaluation (see also Appendix \ref{sec:numerical}). Under our \texttt{pop-cosmos} diffusion model prior, we are able to run MCMC chains (500 iterations; 512 walkers) for a batch of 1,000 galaxies in around 4 hours on a single GPU. This is the largest batch size we are able to sample whilst staying within the 80~GB GPU memory of an NVIDIA A100 card (which we use for many of the computations). This makes posterior sampling feasible for COSMOS-scale datasets of $\gtrsim100,000$ galaxies in around 400 GPU-hrs. The scaling with batch size is non-linear, meaning larger batches can be processed with greater relative efficiency (see Appendix \ref{sec:numerical}). We thus report all compute times based on the optimal configuration we found for a single GPU card. Inference for the full sample was performed by splitting the sample into batches and running these in parallel on multiple GPUs.

For the less memory- and computationally-intensive \texttt{Prospector} prior, the speedup afforded by the \texttt{Speculator} emulator allows full posterior sampling for 4,000 galaxies in 40 minutes on a single GPU\footnote{All run times in the text correspond to NVIDIA Ampere A100 GPUs. We also performed some of our inference using Intel Data Center GPU Max 1550 cards. On this hardware, we found a factor of $2\times$ speedup for inference under the \texttt{Prospector} prior compared to the NVIDIA hardware.}. The \texttt{pop-cosmos} prior is more computationally intensive, as it requires a full ODE solution for every prior evaluation. The throughput for posterior sampling is 250~galaxies per GPU-hr (15 GPU-sec per galaxy) under the \texttt{pop-cosmos} diffusion model prior, and 6,000~galaxies per GPU-hr (0.6~GPU-sec per galaxy) under the \texttt{Prospector} prior. This compares favourably to directly calling \texttt{FSPS} via \texttt{Prospector}, where the throughput for pan-chromatic SED fitting is $\sim\!25$ CPU-hrs {\it per galaxy} \citep{leja19}.

Full Bayesian inference using SPS models of this complexity has been achieved only rarely for catalogs as large as we have analysed here. \citet{leja19} presented one of the largest such analyses, fitting $\sim\!60,000$ galaxies from the 3D-HST survey \citep{skelton14} using a 14 parameter \texttt{Prospector} model\footnote{Other SPS-based analyses of $\mathcal{O}(10^4)$ galaxies have been carried out using the \texttt{BEAGLE} and \texttt{BAGPIPES} models \citep[e.g.][]{chevallard16, carnall18, carnall19}.}. Using the \texttt{dynesty} nested sampler \citep{speagle20} for posterior exploration, \citet{leja19} report a total computation time of $1.5$ million CPU-hrs for the full 3D-HST sample. Subsequent analyses by \citet{leja20, leja22} added a further $\sim\!50,000$ galaxies from COSMOS2015 \citep{laigle16} to their sample. In this work we have been able to perform Bayesian inference on a sample $3\times$ this size, in a relatively modest compute time ($\sim\!1,000$~GPU-hrs in total; 15 GPU-sec per galaxy).

The only other SPS-based analysis of such a large sample is the recent work by \citet{hahn24}, who fit a 13-parameter SPS model \citep{hahn23} to $\sim250,000$ galaxies with spectro-photometry from the DESI Bright Galaxy Survey. To achieve this, \citet{hahn24} use an emulator for \texttt{FSPS} \citep{kwon23}, modelled after the \texttt{Speculator} code \citep{alsing20} that we use here. Combining their emulator with an ensemble slice sampler \citep{karamanis20, karamanis21}, \citet{hahn23, hahn24} report a throughput of 5--10~CPU-min per galaxy.

\section{Conclusions}
\label{sec:conclusions}
The use of a neural network emulator \citep{alsing20} for \texttt{FSPS} \citep{conroy09, conroy10a, conroy10b} has enabled us to perform inference of physical properties and redshifts of individual galaxies using a state-of-the-art 16 parameter SPS model for an unprecedented number ($\sim\!3\times10^5$) of galaxies. We have performed this inference using the empirical priors from \texttt{Prospector} \citep{leja17, leja18, leja19, leja19_sfh, johnson21}, and using a machine-learned population model that was trained on the COSMOS2020 \citep{weaver22} photometric galaxy sample by \citet{alsing24}. In this work, we have applied this methodology to COSMOS2020 data, but the \texttt{pop-cosmos} population distribution can readily applied to \emph{any} survey with comparable depth to COSMOS2020.

On an $r<25$~mag subset (140,745 galaxies) of the COSMOS2020 catalog \citep{weaver22}, inference under the \texttt{pop-cosmos} prior performs favourably compared to both the \texttt{Prospector} prior and photo-$z$ estimates made using \texttt{LePhare} \citep{arnouts99, ilbert06, ilbert09} and \texttt{EAZY} \citep{brammer08} by \citet{weaver22}. For the subset of COSMOS2020 with spectroscopic redshifts (12,014 galaxies), we find that the posterior median redshifts estimated under the \texttt{pop-cosmos} prior show the least bias as a function of magnitude and $z^\text{spec}$, the smallest scatter, and the lowest outlier rate. This holds fairly consistently for a range of $z^\text{spec}$ values and $r$-band magnitudes. Extending to fainter galaxies ($\texttt{irac1}<26$; $25<r\lesssim27$) than were included in the training sample used to calibrate the model ($r<25$), we find that the results under the \texttt{pop-cosmos} prior remain reliable. For the 297 fainter galaxies with measured spectroscopic redshifts, we find that our photo-$z$ estimates are not significantly biased, and that the outlier rate is low ($\sim\!7\%$; compared to $\sim\!9\%$ and $\sim\!16\%$ for \texttt{LePhare} and \texttt{EAZY} respectively).

Our results demonstrate that a flexible physical prescription for galaxy SEDs can be combined with a well-calibrated population model to yield highly accurate photometric redshifts without the use of a spectroscopic training set. The combination of emulators and hardware acceleration straightforwardly enables full Bayesian inference for samples of millions of galaxies. In the large sample limit, we are able to complete MCMC sampling at a rate of 15~GPU-sec per galaxy under the \texttt{pop-cosmos} prior, or 0.6~GPU-sec per galaxy under the \texttt{Prospector} prior. Computation is batched, allowing batches of $\mathcal{O}(10^3)$ galaxies to be sampled in parallel on multiple separate GPU cards.

In addition to redshift, we have access to a posterior distribution over 15 other physical parameters for each galaxy, enabling rich astrophysical insights in tandem with cosmological analysis. In ongoing work, we are applying the tools developed here and in \citet{alsing24} to current datasets -- for both cosmology (Loureiro et al.\ in prep.) and galaxy evolution science (Deger et al.\ in prep.). This work therefore opens up the pathway to leverage these methods on Stage IV surveys such as LSST, {Roman}, and {\it Euclid}.

\vfill
\noindent\textbf{Author contributions.} 
We outline the different contributions below using keywords based on the CRediT (Contribution Roles Taxonomy) system. %
\textbf{ST:} conceptualization, methodology, software, formal analysis, visualization, writing -- original draft, writing -- editing \& review.
\textbf{JA:} conceptualization, methodology, software, validation, writing -- editing \& review.
\textbf{HVP:} conceptualization, validation, writing -- editing \& review, supervision, funding acquisition.
\textbf{SD:} software, visualization, validation, writing -- editing \& review.
\textbf{DJM:} methodology, validation, writing -- editing \& review.
{\bf BL:} data curation, validation, writing -- editing \& review.
\textbf{JL:} validation, writing -- editing \& review.
\textbf{AL:} writing -- editing \& review.

\noindent\textbf{Data availability.}
A catalog of photometric redshifts is available on \dataset[Zenodo]{https://doi.org/10.5281/zenodo.13627488}. The COSMOS catalog \citep{weaver22} is publicly available at the\dataset[ESO Archive]{https://doi.org/10.18727/archive/52} and the\dataset[COSMOS2020 webpage]{https://cosmos2020.calet.org}. \add{The C3R2 spectroscopic catalog \citep{masters17} is hosted at the \dataset[Keck Observatory Archive]{https://koa.ipac.caltech.edu/Datasets/C3R2/}. The FMOS--COSMOS spectroscopic catalog \citep{kashino19} is hosted by the \dataset[Kavli IMPU]{http://member.ipmu.jp/fmos-cosmos/FMOS-COSMOS.html}.}

\noindent\textbf{Acknowledgments.} 
We thank John Weaver for helpful and prompt answers to queries about COSMOS2020. We thank Katherine Kauma, Priscila Pessi, \add{and Joel Johansson} for useful discussions. \add{We thank the referee for helpful feedback which has improved the paper. ST thanks the Strasbourg astronomical Data Centre (CDS) team at EAS2024 for useful advice about Aladin.} ST, JA, HVP and SD have been supported by funding from the European Research Council (ERC) under the European Union's Horizon 2020 research and innovation programmes (grant agreement no. 101018897 CosmicExplorer). This work has been enabled by support from the research project grant ‘Understanding the Dynamic Universe’ funded by the Knut and Alice Wallenberg Foundation under Dnr KAW 2018.0067. HVP was additionally supported by the G\"{o}ran Gustafsson Foundation for Research in Natural Sciences and Medicine. BL is supported by the Royal Society through a University Research Fellowship. 

%


\facilities{This study utilizes observations collected at the European Southern Observatory under ESO programme ID 179.A-2005 and 198.A-2003 and on data products produced by CALET and the Cambridge Astronomy Survey Unit on behalf of the UltraVISTA consortium. \add{This research has made use of the Keck Observatory Archive (KOA), which is operated by the W.~M.\ Keck Observatory and the NASA Exoplanet Science Institute (NExScI), under contract with the National Aeronautics and Space Administration.} This research utilized the Sunrise HPC facility supported by the Technical Division at the Department of Physics, Stockholm University. This work was performed using resources provided by the Cambridge Service for Data Driven Discovery (CSD3) operated by the University of Cambridge Research Computing Service (\url{www.csd3.cam.ac.uk}), provided by Dell EMC and Intel using Tier-2 funding from the Engineering and Physical Sciences Research Council (capital grant EP/T022159/1), and DiRAC funding from the Science and Technology Facilities Council (\url{www.dirac.ac.uk}). This work was performed using resources provided by Cambridge Dawn AI Service as part of the UK AI Research Resource operated by the University of Cambridge Research Computing Service (\url{www.hpc.cam.ac.uk}), funded by UKRI, DSIT, Dell EMC and Intel.}

\software{\texttt{NumPy} \citep{harris20}; \texttt{SciPy} \citep{virtanen20}; \texttt{Matplotlib} \citep{hunter07}; \texttt{corner} \citep{dfm16}; \texttt{PyTorch} \citep{paszke19}; \texttt{Speculator} \citep{alsing20}; \texttt{torchdiffeq} \citep{chen18}; \texttt{prospector} \citep{johnson21}; \texttt{FSPS} \citep{conroy09, conroy10a, conroy10b}; \texttt{python-fsps} \citep{pythonfsps}; \texttt{tqdm} \citep{tqdm}; \texttt{affine}\footnote{\url{https://github.com/justinalsing/affine}}; \texttt{ffjord-lite}\footnote{\url{https://github.com/jackgoffinet/ffjord-lite}}; \texttt{intel-extension-for-pytorch}\footnote{\url{https://github.com/intel/intel-extension-for-pytorch}}. We have made use of color schemes prepared by \citet{tol21} and \citet{green11}. \add{This research has made use of the Aladin Sky Atlas \citep{bonnarel00, boch14, baumann22} developed at CDS, Strasbourg Observatory, France.}}
%

%
\appendix
\section{Numerical Experiments}
\label{sec:numerical}
In this appendix, we report the results of numerical experiments relating to the accuracy of the ODE solver used in computing the log prior under our \texttt{pop-cosmos} diffusion model (see Equations \ref{eq:psolve} and \ref{eq:xsolve}). \citet{grathwohl18} also conduct numerical tests on a one-dimensional density estimation problem, finding numerical error to be negligible for solver tolerances of $\leq10^{-5}$. In our fiducial results, we set both the absolute and relative error tolerances of the \citet{dormand80} Runge-Kutta solver to $10^{-4}$, for reasons of computational efficiency. We found almost no change in actual photometric redshift performance (as measured by the metrics in Figure \ref{fig:photo-z-metrics}) between using a tolerance of $10^{-4}$, an even less strict tolerance of $10^{-3}$, or a stricter tolerance of $10^{-5}$. So, we expect that this choice has a relatively small impact on the quality of photo-$z$ estimates under our model.

As a further test, we take $256,000$ draws of our SPS parameters ($\bm{\vartheta}$) from our \texttt{pop-cosmos} prior. For our standard MCMC configuration (see Section \ref{sec:mcmc}), this is the number of log posterior evaluations required in a single step of all 256 ensemble walkers for $1,000$ galaxies. We then use Equation \eqref{eq:logprob} to evaluate the log probability $\ln P(\bm{\vartheta})$ of these draws at a variety of ODE solver tolerances between $10^{-3}$ and $10^{-9}$. 

\begin{figure}
    \centering
    \includegraphics[width=\linewidth]{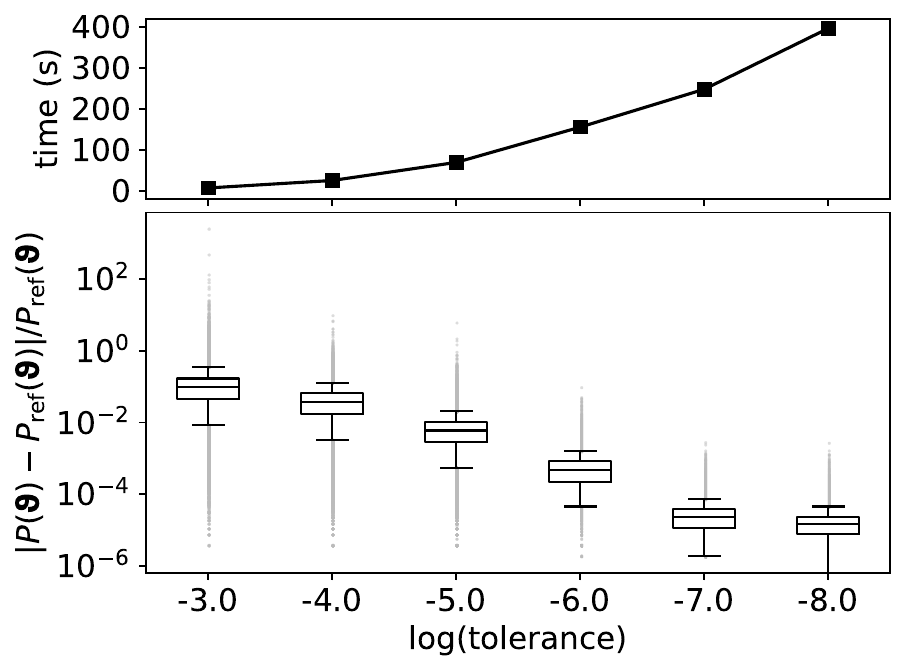}
    \caption{Box-and-whisker plot showing the distribution of fractional errors made in $P(\bm{\vartheta})$ for $256,000$ different $\bm{\vartheta}$, using different ODE solver tolerances (measured relative to the reference value of $P_\text{ref}(\bm{\vartheta})$ computed with $\texttt{atol}=\texttt{rtol}=10^{-9}$). Boxes show the median and interquartile range. Whiskers show the 5th and 95th percentiles. Outliers are shown as grey points. Upper panel shows the time taken for 256,000 prior evaluations on a single NVIDIA A100 GPU.}
    \label{fig:error-vs-tol}
\end{figure}

\begin{figure}
    \centering
    \includegraphics[width=\linewidth]{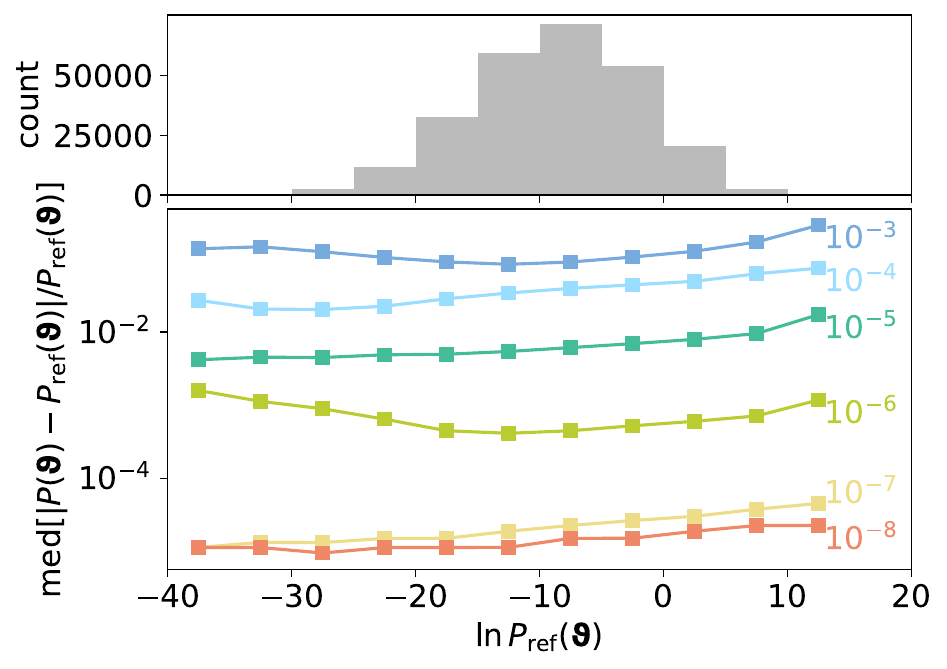}
    \caption{Median size of the fractional error on $P(\bm{\vartheta})$, binned by $\ln P_\text{ref}(\bm{\vartheta})$ for a range of solver tolerances.}
    \label{fig:error-vs-lnp}
\end{figure}

Figure \ref{fig:error-vs-tol} shows the distribution of fractional errors made relative to $P_\text{ref}(\bm{\vartheta})$ computed with the most stringent tolerance setting of $10^{-9}$. We also show the time taken to carry out the 256,000 $\ln P(\bm{\vartheta})$ evaluations on a single NVIDIA A100 GPU. For our fiducial tolerance of $10^{-4}$, the median size of the fractional error on $P(\bm{\vartheta})$ is $3.7\%$. The interquartile range spans between $1.7$ and $6.6$\%, and the size of the fractional error is below the 10\% level $\sim\!90\%$ of the time. The execution time with tolerance of $10^{-4}$ was $25.1$~s for the full set of 256,000 $\bm{\vartheta}$ draws. For a tolerance of $10^{-5}$, this increases to $69.2$~s. The median size of the fractional error shifts to around $0.6\%$, with $>99\%$ of cases having a fractional error below 10\%, and 95\% having a fractional error below $2.1\%$. In Figure \ref{fig:error-vs-lnp}, we plot the median size of the fractional error, binned by $\ln P_\text{ref}(\bm{\vartheta})$. For all solver tolerances tested, the median size of the fractional error tends to be highest towards regions of very high prior density.

\begin{figure}
    \centering
    \includegraphics[width=\linewidth]{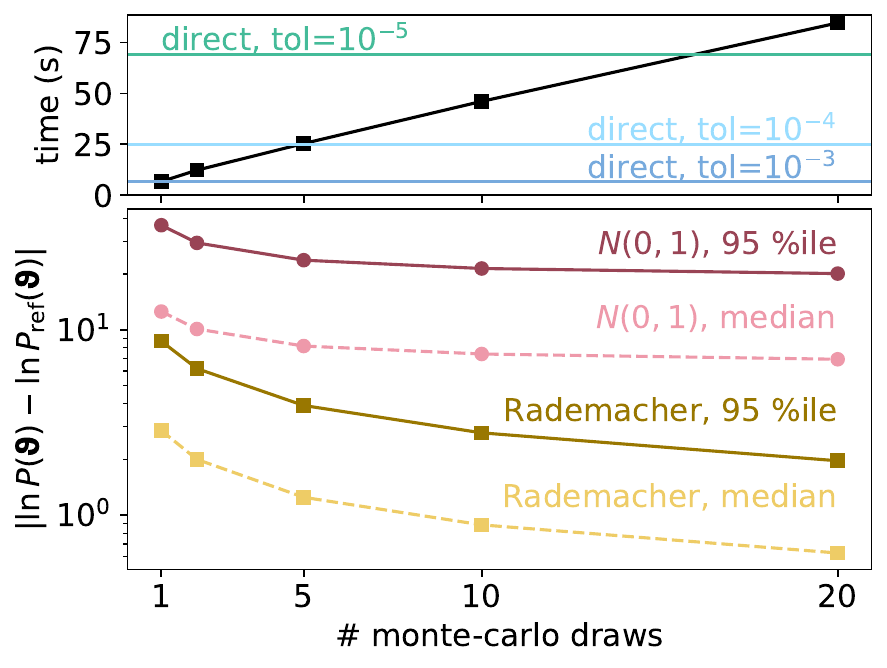}
    \caption{Absolute error in $\ln P(\bm{\vartheta})$ for different configurations of the Hutchinson trace estimator. Deviations are computed relative to a directly computed $\ln P_\text{ref}(\bm{\vartheta})$ with $\texttt{atol}=\texttt{rtol}=10^{-5}$. Upper panel shows the time taken for 256,000 prior evaluations using each configuration. Horizontal lines show the direct solution with different tolerances.}
    \label{fig:error-vs-mc}
\end{figure}

\begin{figure}[t!]
    \centering
    \includegraphics[width=\linewidth]{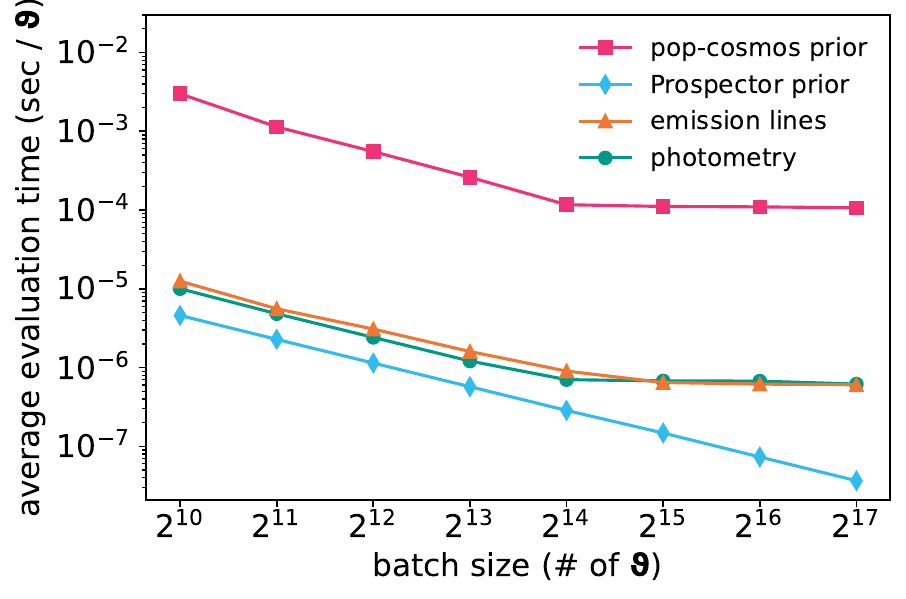}
    \caption{Average function evaluation time vs.\ batch size (i.e.\ number of $\bm{\vartheta}$ draws passed to the function) for the dominant functions in our log-posterior evaluation. Times are in GPU-sec and correspond to an NVIDIA A100 GPU.}
    \label{fig:scaling}
\end{figure}

We also test the behaviour of the \citet{hutchinson89} trace estimator given in Equation \eqref{eq:hutch}. For these tests, we use the direct log prior evaluation with $\texttt{atol}=\texttt{rtol}=10^{-5}$ as our reference $\ln P_\text{ref}(\bm{\vartheta})$. We perform tests with the noise vector $\bm{\epsilon}$ in the vector--Jacobian product drawn from either a Rademacher distribution or a unit normal, with 1--20 Monte-Carlo samples being used to compute the expectation in Equation \eqref{eq:hutch}. Figure \ref{fig:error-vs-mc} shows the median absolute deviation, and the 95th percentile of $|\ln P(\bm{\vartheta}) - \ln P_\text{ref}(\bm{\vartheta})|$ for the different configurations tested. We find that the estimator with $\bm{\epsilon}$ drawn from a Rademacher distribution converges faster towards the direct result (see also extensive discussion in \citealp{avron11}). The execution time is approximately linear with the number of Monte-Carlo samples. After 20 Monte-Carlo repeats, the computation time has exceeded that of the direct calculation with the same tolerance. From this, we conclude that the direct solution is preferable in our pipeline.

Finally, we test how the GPU evaluation time of the three dominant terms in the posterior scales with batch size. To test this, we begin by taking $n=2^{10},\dots,2^{17}$ draws of $\bm{\vartheta}$ from our prior. For each $n$ and corresponding set of $\bm{\vartheta}$ draws, we measure the time taken to evaluate the \texttt{pop-cosmos} prior, evaluate the \texttt{Prospector} prior, emulate 26-band COSMOS photometry using \texttt{Speculator}, and emulate the contributions of the 44 emission lines. From this, we calculate the effective time taken per function evaluation for the four functions. Figure \ref{fig:scaling} shows our results. The \texttt{pop-cosmos} prior was evaluated with $\texttt{atol}=\texttt{rtol}=10^{-4}$ as in our fiducial configuration. We can see that all of the functions saturate around $n=2^{14}=16,384$, except for the \texttt{Prospector} prior which is much cheaper to compute and only begins to saturate by $n=2^{21}=2,097,152$. The saturation at $2^{14}$ means that MCMC sampling on batches of $\lesssim32$ galaxies (assuming 512 walkers, as per \S\ref{sec:mcmc}), or MAP estimation on batches of $\lesssim16,384$ galaxies will not be optimally efficient. The \texttt{pop-cosmos} prior saturates at an average evaluation time of $0.1$~ms, as expected from the results earlier in this appendix. For all batch sizes, calls to the photometry and emission line emulators are a factor of $\gtrsim100\times$ cheaper than diffusion model calls. For batch sizes $>2^{14}$, the cost of evaluating the \texttt{Prospector} prior is effectively negligible in comparison to the emulator calls.

\section{\add{Example Spectra}}
\label{sec:spectra}

\add{Two of the galaxies shown in Figures \ref{fig:sedfits} and \ref{fig:redshiftposteriors} have highly discrepant photometric redshift estimates, differening from the spectroscopic values by $|\Delta_z|>0.15$.  We examine their spectra more closely here.}

\begin{figure*}
    \centering
    \includegraphics[width=\linewidth]{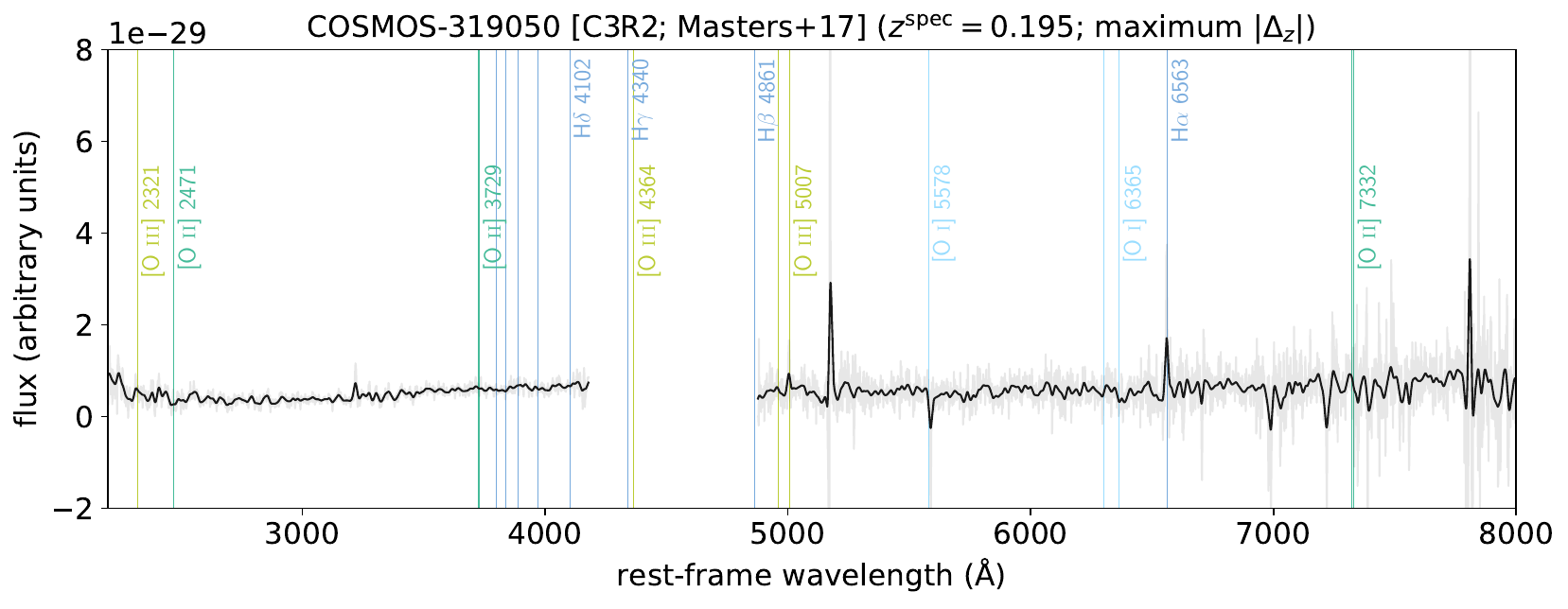}
    \includegraphics[width=\linewidth]{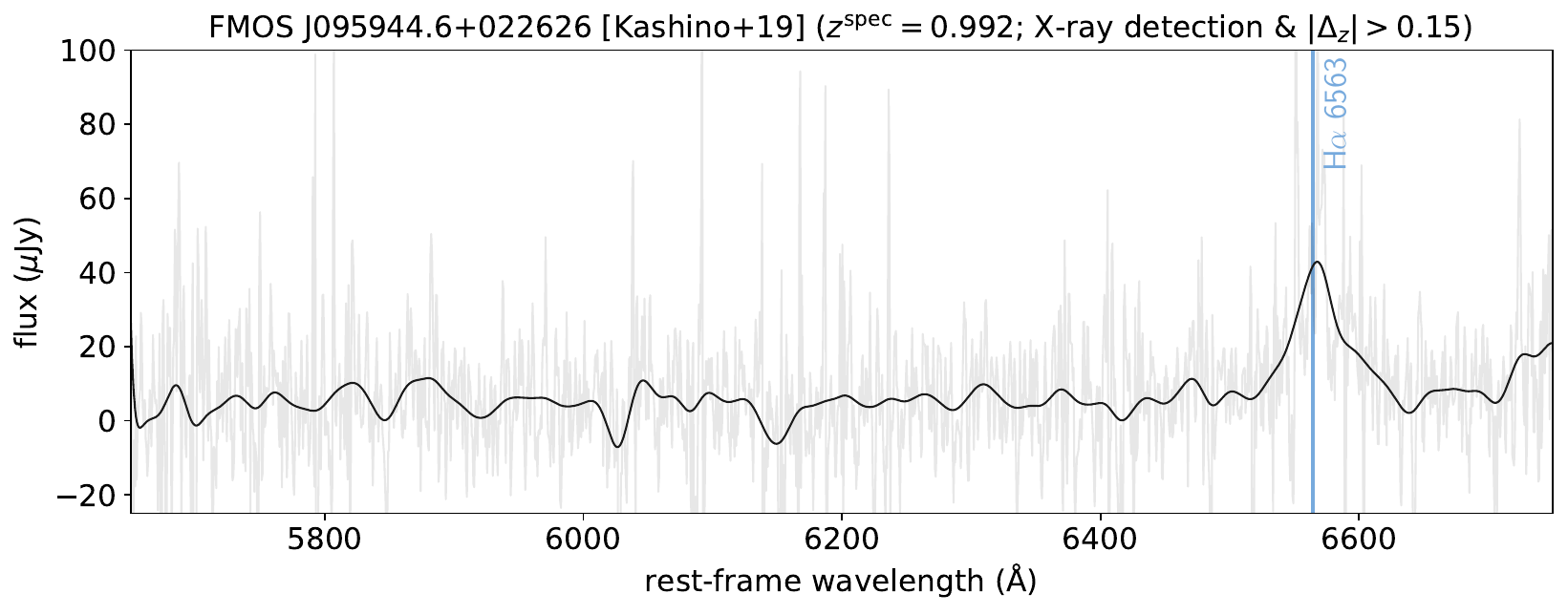}
    \caption{\add{Spectra corresponding to the two fits on the right hand side of Figures \ref{fig:sedfits} and \ref{fig:redshiftposteriors}. The faint grey lines are the raw spectra, and the solid black lines are smoothed with a Gaussian kernel. Spectra are de-redshifted into the rest-frame based on the reported $z^\text{spec}$. Central wavelengths of Balmer lines are overplotted in both panels. Oxygen emission lines are overplotted in the top panel only. \textbf{Top panel:} spectrum for the galaxy with the largest $|\Delta_z|$ (top right of Figs.\ \ref{fig:sedfits} and \ref{fig:redshiftposteriors}). \textbf{Bottom panel:} spectrum for the X-ray detected example galaxy with $|\Delta_z|>0.15$ (bottom right of Figs.\ \ref{fig:sedfits} and \ref{fig:redshiftposteriors}).}}
    \label{fig:spectra}
\end{figure*}

\add{The top panel of Figure \ref{fig:spectra} shows the Keck Low Resolution Imaging Spectrometer (LRIS; \citealp{oke95}) spectrum for COSMOS-319050 from \citet{masters17}. This is the galaxy with maximum $|\Delta_z|$  (upper right panel of Figures \ref{fig:sedfits} and \ref{fig:redshiftposteriors}). The spectrum shows a flat continuum with several prominent emission lines. We show the spectrum de-redshifted by the reported $z^\text{spec}=0.195$. The original spectrum is shown with a faint grey line, and a smoothed\footnote{\add{We apply a Gaussian kernel smoother with a bandwidth of 6~\AA\ using the Nadaraya--Watson estimator \citep{nadaraya64, watson64}.}} version is shown as a solid black line. We overplot the rest frame wavelengths (taken from \citealp{byler17}) of the Balmer series, and of any oxygen lines within the wavelength range covered. There are visible lines that coincide with H$\alpha$ 6563~\AA\ and [O$\,\textsc{iii}$] 5007~\AA\ (H$\beta$ cannot be identified as it falls between the red and blue channels of LRIS). There are no clear line matches at $z\approx2.64$. Assuming $z=0.195$, the strongest apparent line in the spectrum is at 5170~\AA, which does not match any of the 128 nebular emission lines used in \citet{byler17}. The galaxy is faint and red, with $r=24.86$ and $i=24.89$ in the COSMOS2020 catalog, and is not visible in the majority of legacy imaging available in Aladin \citep{bonnarel00}.}  

\add{The galaxy in the lower right panel of Figures \ref{fig:sedfits} and \ref{fig:redshiftposteriors} (FMOS J095944.6+022626; coordinates 09h\,59m\,44.61s, +02$^\circ$\,26$'$\,26.13$''$) has spectroscopic data from the Subaru FMOS survey \citep{kashino19}. This source has an X-ray detection in the Chandra COSMOS Legacy catalog (ID 1186; \citealp{civano16}). The spectrum for this galaxy is shown in the lower panel of Figure \ref{fig:spectra}, de-redshifted by the reported $z^\text{spec}=0.992$. In the smoothed spectrum\footnote{\add{Smoothed with a bandwidth of 8~\AA.}}, a broad H$\alpha$ line is the most distinct feature, characteristic of an AGN \citep[see e.g.][]{schulze18}. In the observer frame this line would fall at 13073~\AA, coinciding with the $J$-band.}

%
%
\bibliography{sample631}
\bibliographystyle{aasjournal}
%


\end{document}